\documentclass[twocolumn,trackchanges]{aastex631}

\shorttitle{Star Formation Activity Beyond the Outer Arm II}
\shortauthors{Izumi et al.}
\graphicspath{{./}{figures/}}

\begin{document}

\title{Star Formation Activity Beyond the Outer Arm II: Distribution and Properties of Star Formation}

\author[0000-0003-1604-9127]{Natsuko Izumi}
\affiliation{Institute of Astronomy and Astrophysics, Academia Sinica, No. 1, Section 4, Roosevelt Road, Taipei10617, Taiwan}

\author{Naoto Kobayashi}
\affiliation{Laboratory of Infrared High-resolution spectroscopy (LIH), Koyama Astronomical Observatory, Kyoto Sangyo University, Motoyama, Kamigamo, Kita-ku, Kyoto 603-8555, Japan}
\affiliation{Institute of Astronomy, School of Science, University of Tokyo, 2-21-1, Osawa, Mitaka, Tokyo 181-0015, Japan}
\affiliation{Kiso Observatory, Institute of Astronomy, School of Science, University of Tokyo,10762-30 Mitake, Kiso-machi, Kiso-gun, Nagano 397-0101, Japan}

\author{Chikako Yasui}
\affiliation{National Astronomical Observatory of Japan, 2-21-1, Osawa, Mitaka, Tokyo, 181-8588, Japan}

\author{Masao Saito}
\affiliation{National Astronomical Observatory of Japan, 2-21-1, Osawa, Mitaka, Tokyo, 181-8588, Japan}
\affiliation{The Graduate University of Advanced Studies, (SOKENDAI), 2-21-1 Osawa, Mitaka, Tokyo 181-8588, Japan}

\author{Satoshi Hamano}
\affiliation{National Astronomical Observatory of Japan, 2-21-1, Osawa, Mitaka, Tokyo, 181-8588, Japan}

\author{Patrick M. Koch}
\affiliation{Institute of Astronomy and Astrophysics, Academia Sinica, No. 1, Section 4, Roosevelt Road, Taipei10617, Taiwan}



\begin{abstract}

The outer Galaxy beyond the Outer Arm represents a promising opportunity to study star formation in an environment vastly different from the solar neighborhood. 
In our previous study, we identified 788 candidate star-forming regions in the outer Galaxy (at galactocentric radii $R_{\rm G}$ $\ge$ 13.5 kpc) based on Wide-field Infrared Survey Explorer (WISE) mid-infrared (MIR) all-sky survey.
In this paper, we investigate the statistical properties of the candidates and their parental molecular clouds derived from the Five College Radio Astronomy Observatory (FCRAO) CO survey. 
We show that the molecular clouds with candidates have a shallower slope of cloud mass function, a larger fraction of clouds bound by self-gravity, and a larger density than the molecular clouds without candidates.
To investigate the star formation efficiency (SFE) at different $R_{\rm G}$, we used two parameters: 1) the fraction of molecular clouds with candidates and 2) the monochromatic MIR luminosities of candidates per parental molecular cloud mass.
We did not find any clear correlation between SFE parameters and $R_{\rm G}$ at $R_{\rm G}$ of 13.5 kpc to 20.0 kpc, suggesting that the SFE is independent of environmental parameters such as metallicity and gas surface density, which vary considerably with $R_{\rm G}$.
Previous studies reported that the SFE per year (SFE/yr) derived from the star-formation rate surface density per
total gas surface density, \ion{H}{1} plus H$_2$, decreases with increased $R_{\rm G}$.
Our results might suggest that the decreasing trend is due to a decrease in \ion{H}{1} gas conversion to H$_2$ gas.

\end{abstract}



\section{Introdution}\label{sec:1}
The outer Galaxy beyond the Outer Arm provides an excellent opportunity to study star formation in an environment significantly different from the solar neighborhood.
For example, gas density and metallicity in the outer Galaxy are lower than in the solar neighborhood \citep[e.g.,][]{Smartt1997, Wolfire2003, Fernandez2017}.
Furthermore, the interstellar medium (ISM) is dominated by \ion{H}{1}, so that H$_2$ fractions are extremely small \citep[e.g.,][]{Wolfire2003}.
From the radial profile of gas surface densities in our Galaxy \citep[e.g.][]{Heyer2015,Nakanishi2016},
the entire (\ion{H}{1}+H$_2$) and \ion{H}{1} gas surface densities ($\Sigma_{\rm gas}$ and $\Sigma_{\rm HI}$, respectively)
start to decrease at the galactocentric radii ($R_{\rm G}$) of about 13.5 kpc.
Both values at $R_{\rm G}$ of about 20 kpc are about one-tenth of those in the solar neighborhood.
The H$_2$ gas surface density ($\Sigma_{\rm H_2}$) at $R_{\rm G}$ of 13.5 and 20 kpc are
about one-fifth and less than one-tenth of that in the solar neighborhood, respectively.
From the radial profile of the metallicity \citep[e.g.,][]{Smartt1997,Fernandez2017},
the metallicity at $R_{\rm G}$ of 13.5 and 20 kpc are about half and less than one-fifth of that in the solar neighborhood, respectively.
Such environments may have similar characteristics as dwarf galaxies and our Galaxy in the early phase of the formation, particularly in the thick disk formation 
\citep{Ferguson1998a,Kobayashi2008}.
Therefore, we can directly observe galaxy formation processes in unprecedented detail at a much closer distance than distant galaxies.

In low-gas-density environments, the star-formation rate (SFR) is known to decrease significantly.
It is shown in the empirical relation between SFR surface density ($\Sigma_{\rm SFR}$) and $\Sigma_{\rm gas}$ in nearby galaxies known as the Kennicutt-Schmidt law
\citep[$\Sigma_{\rm SFR}$ $\propto$ $\Sigma_{\rm gas}^n$;][]{Schmidt1959,Kennicutt1998,Kennicutt2012}.
The exponent $n$ in the Kennicutt-Schmidt law increases from the usual $n$ $\sim$ 1
(10 $M_\odot$pc$^{-2}$ $\lesssim$ $\Sigma _{\rm gas}$ $\lesssim$ 100 $M_\odot$pc$^{-2}$) 
to $n$ $\sim$ 2 in regions that have a surface density less than 10 $M_\odot$pc$^{-2}$
\citep[e.g., Figure 15 in ][]{Bigiel2008}.
This implies that the star-formation efficiency per year (SFE/yr) derived from $\Sigma_{\rm SFR}$/$\Sigma _{\rm gas}$
is constant over $\Sigma_{\rm gas}$ in regions with relatively high gas surface densities (10 $M_\odot$pc$^{-2}$ $\lesssim$ $\Sigma _{\rm gas}$ $\lesssim$ 100 $M_\odot$pc$^{-2}$),
but it decreases with decreasing
$\Sigma_{\rm gas}$ in regions with lower gas surface densities ($\Sigma _{\rm gas}$ $\lesssim$ 10 $M_\odot$pc$^{-2}$).
\citet{Shi2014} reported that the SFE/yr of the nearby dwarf galaxies is about one-tenth of that in spiral galaxies with similar gas densities.
Metallicities of those dwarf galaxies are about one-tenth of that in the solar neighborhood.
However, the mechanisms for this qualitative change of star formation have not been understood because detailed observations of star-forming regions have been impossible, even in nearby galaxies. 

In our Galaxy, the SFE/yr ($\Sigma_{\rm SFR}$/$\Sigma _{\rm gas}$) is derived up to $R_{\rm G}$ of $\sim$ 15 kpc \citep{Kennicutt2012}.
The SFE/yr is constant over $R_{\rm G}$ for $R_{\rm G}$ $\lesssim$ 13.5 kpc (10 $M_\odot$pc$^{-2}$ $\lesssim$ $\Sigma _{\rm gas}$), but then decreases for $R_{\rm G}$ $>$ 13.5 kpc ($\Sigma _{\rm gas}$ $\lesssim$ 10 $M_\odot$pc$^{-2}$).
The SFE/yr at 15 kpc is about one-fourth of that in the solar neighborhood.
These trends are consistent with those reported in the nearby galaxies.
The outer Galaxy is much closer than any galaxy, and therefore, the most detailed study of the star-forming region is possible. 

To investigate the global properties of star formation activities in the outer Galaxy,
we developed a simple criterion to identify unresolved distant young star-forming regions.
The criterion is  based on the color-color diagram of mid-infrared (MIR) all-sky survey data by \textit{Wide-field Infrared Survey Explorer (WISE)}
\citep{Wright2010,Jarrett2011}: [3.4] - [4.6] $\ge$ 0.5, [4.6] - [12] $\ge$ 2.0, and [4.6] - [12] $\le$ 6.0
\citep[][hereafter Paper I]{Izumi2017}.
Furthermore, young star-forming regions (age $<$ 3 Myr) should accompany their parental molecular clouds \citep{Lada2003}.
Therefore, the criterion enables us to pick up star-forming regions effectively by combining them with CO survey data in the outer Galaxy. 
In paper I, we applied the criterion to 466 molecular clouds in the outer Galaxy at
$R_{\rm G}$ of more than 13.5 kpc
detected from the Five College Radio Astronomy Observatory (FCRAO)
$^{12}$CO(1-0) survey of the outer Galaxy \citep[survey region: 102$^\circ$.49 $\le$ $l$ $\le$ 141$^\circ$.54, 3$^\circ$.03 $\le$ $b$ $\le$ 5$^\circ$.41, -153 km s$^{-1}$ $\le$ $v_{\rm LSR}$ $\le$ +40 km s$^{-1}$;][]{Heyer1998, Brunt2003}.
As a result, we identified 788 \textit{WISE} sources in 252 molecular clouds as candidate star-forming regions at $R_{\rm G}$ of up to $\sim$ 20 kpc.
Among the 788 \textit{WISE} sources, 77 \textit{WISE} sources are already detected as star-forming regions
from previous near-infrared (NIR) observations \citep[e.g.,][]{Snell2002}.
Therefore, our survey newly identified 711 \textit{WISE} sources in 240 molecular clouds as candidate star-forming regions. 

Additionally, we investigated the possible contamination for identified candidate star-forming regions by foreground and/or background objects. 
To estimate the contamination rate in these candidates quantitatively, we compared the number density of the \textit{WISE} sources, which satisfies the criterion, in the cloud region with those in the field region.
Based on the number distribution of the contamination rate, we set the upper contamination threshold to 30 \% for a candidate to be regarded as reliable  (see details in Section 4.2 in Paper I). 
Among all the 252 molecular clouds,  211 molecular clouds satisfy this condition.

In this paper, we discuss the statistical properties of the candidate star-forming regions and their parental molecular clouds identified in Paper I.
Section 2 describes the FCRAO CO data and \textit{WISE} MIR data, including sensitivities at different $R_{\rm G}$. 
Sections 3 and 4 investigate the properties of molecular clouds (with and without candidates) and candidates themselves, respectively. 
In Section 5, we discuss the variation of star-formation activities at different $R_{\rm G}$.
We conclude the paper in Section 6.
More detailed information is given in the appendices for the threshold values of the FCRAO and \textit{WISE} data (Appendix \ref{sec:a_1}) and the distribution of molecular clouds (Appendix \ref{sec:a_2}).

\section{Data}\label{sec:2}
\subsection{FCRAO CO data}\label{sec:2_1}
To investigate the properties of molecular clouds detected by the FCRAO survey, we employed the molecular clouds catalog by
\citet{Brunt2003} (hereafter BKP catalog) in Paper I and in this paper.
In the BKP catalog, the typical sensitivity ($\sigma$) is 0.17 K in the $T_{\rm B}$ scale with a
spatial and velocity resolution of 100$^{\prime \prime}$ and 0.98 km s$^{-1}$, respectively.

\subsubsection{BKP catalog}\label{sec:2_11}
The BKP catalog was generated in a two-phase object identification procedure.
The first phase consists of grouping pixels into contiguous structures above a radiation temperature threshold of 0.8 K ($=$ 4.7$\sigma$).
The second phase decomposes the first-phase objects by an enhanced version of the CLUMPFIND algorithm, using dynamic thresholding, with a threshold of 0.8 K used for discrimination.
A two-dimensional elliptical Gaussian is fitted to the velocity-integrated map of each cloud,
and the resulting major and minor axes and position angles are included in the BKP catalog.
A Gaussian profile is fitted to measure the global linewidth to each cloud's spatially integrated emission line.
Model Gaussian clouds, truncated at 0.8 K, are examined to determine the effects of biases on measured quantities induced by truncation.
Through a detailed analysis of the cataloged clouds, statistical corrections for the effects of truncation on measured sizes, linewidths, and integrated intensities are derived and applied, along with corrections for the effects of finite resolution on the measured attributes \citep[see details in][]{Brunt2003}.

To estimate the properties of molecular clouds (hereafter BKP clouds), we used the basic attributes of the BKP clouds tabulated in the BKP catalog:
1) Galactic coordinates ($l$, $b$),
2) Local Standard of Rest (LSR) velocity ($v_{\rm LSR}$),
3) number of spatial pixels over which the velocity is integrated for a cloud ($N_{\rm S}$),
4) major ($l_{\rm major}$) and minor ($l_{\rm minor}$) diameter (full-width half-maximum; FWHM) derived from a two-dimensional elliptical Gaussian fit,
5) FWHM linewidth ($dv$) derived from Gaussian fit,
and 6) integrated CO line intensity ($I_{\rm CO}$ =  $\int T_{\rm B} dv$).

\citet{Brunt2003} applied the statistical correction for only clouds with a peak temperature of more than 1.6 K (two times the threshold value).
Among all 466 clouds, corrected $l_{\rm major}$ and $l_{\rm minor}$ are derived only for 166 clouds.
Corrected $dv$ and $I_{\rm CO}$ are derived only for 139 and 176 clouds, respectively.
To estimate the properties for all 466 clouds, we used raw-fitted (not corrected) values for the rest of the clouds.
We note that the difference between corrected value and raw-fitted value is less than a factor of two. 

\begin{figure*}
\epsscale{1.0}
\plotone{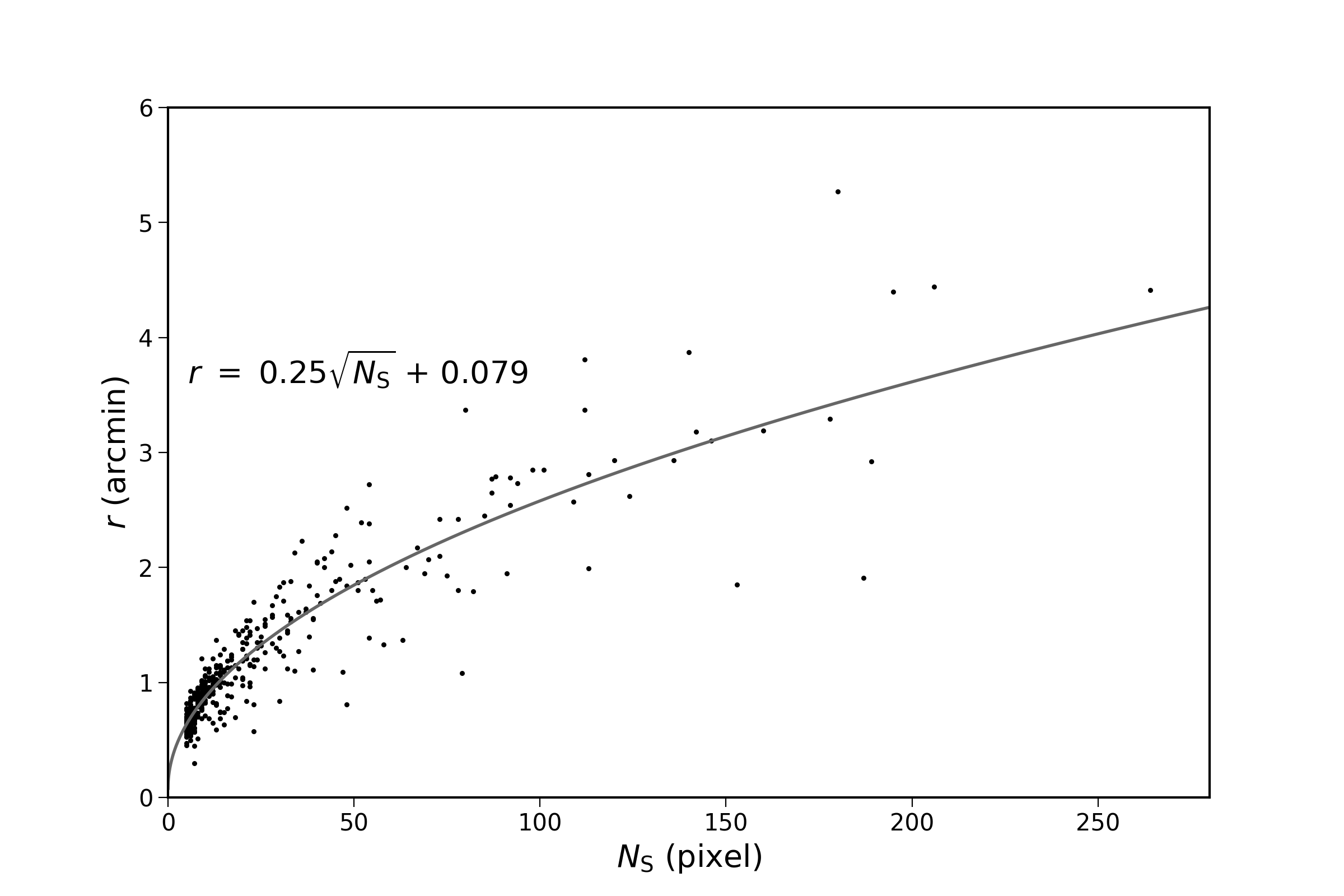}
\caption{
Relation between number of spatial pixels in clouds ($N_{\rm S}$) and radius ($r$) of clouds in the BKP catalog.
The gray curve shows the result of the least-squares fitting ($r$ = 0.25 $\sqrt{N_{\rm S}}$ + 0.079) for clouds in the range of 5 $\le$ $N_{\rm S}$ $\le$ 100.
}  
\label{r-pix} 
\end{figure*}

Using the galactic coordinates ($l$,$b$) and $v_{\rm LSR}$, we derived the kinematic distance ($D$) of the BKP clouds,
assuming a flat rotation curve with a rotation speed of 220 km s$^{-1}$ and a galactocentric distance of the Sun of 8.5 kpc \citep[e.g.,][]{Heyer2001}.
Cloud radii are not listed in the BKP catalog.
Therefore, we estimated the radius ($r$) from $l_{\rm major}$ and $l_{\rm minor}$:
$r = 0.5 \times \sqrt{(l_{\rm major} \times l_{\rm minor})}$.
The Gaussian fits were attempted for all clouds with $N_{\rm S}$ of larger than 5 pixels.
Among all the 466 clouds, the fitting failed to converge or was not attempted for 86 clouds.
For those cloudlets, we estimated $r$ from $N_{\rm S}$ using the $r$--$N_{\rm S}$ relation estimated by the least-squares fitting for clouds in the range of 5 $\le$ $N_{\rm S}$ $\le$ 100 (Figure \ref{r-pix}).
In this fitting range,
the variation in $r$ is smaller than 2$^\prime$,
and the total number of clouds is 359, which is 94\% of the 380 clouds (= 466 $-$ 86).
The result of the fitting is $r$ = 0.25 ($\pm$ 0.0065) $\sqrt{N_{\rm S}}$ + 0.079 ($\pm$ 0.0029) (gray curve in Figure \ref{r-pix}).
In Figure \ref{r-pix}, the fitted curve appears to be good in the whole fitting range. 
The size at $N_{\rm S}$ = 0 of the fitting curve is only 0.$^\prime$079, which is much smaller than the pixel scale (0.$^\prime$837), thus negligible.
This assures the validity of the fitting.

\subsubsection{Sensitivity of FCRAO data}\label{sec:2_12}
\begin{figure*}
\epsscale{1.0}
\plotone{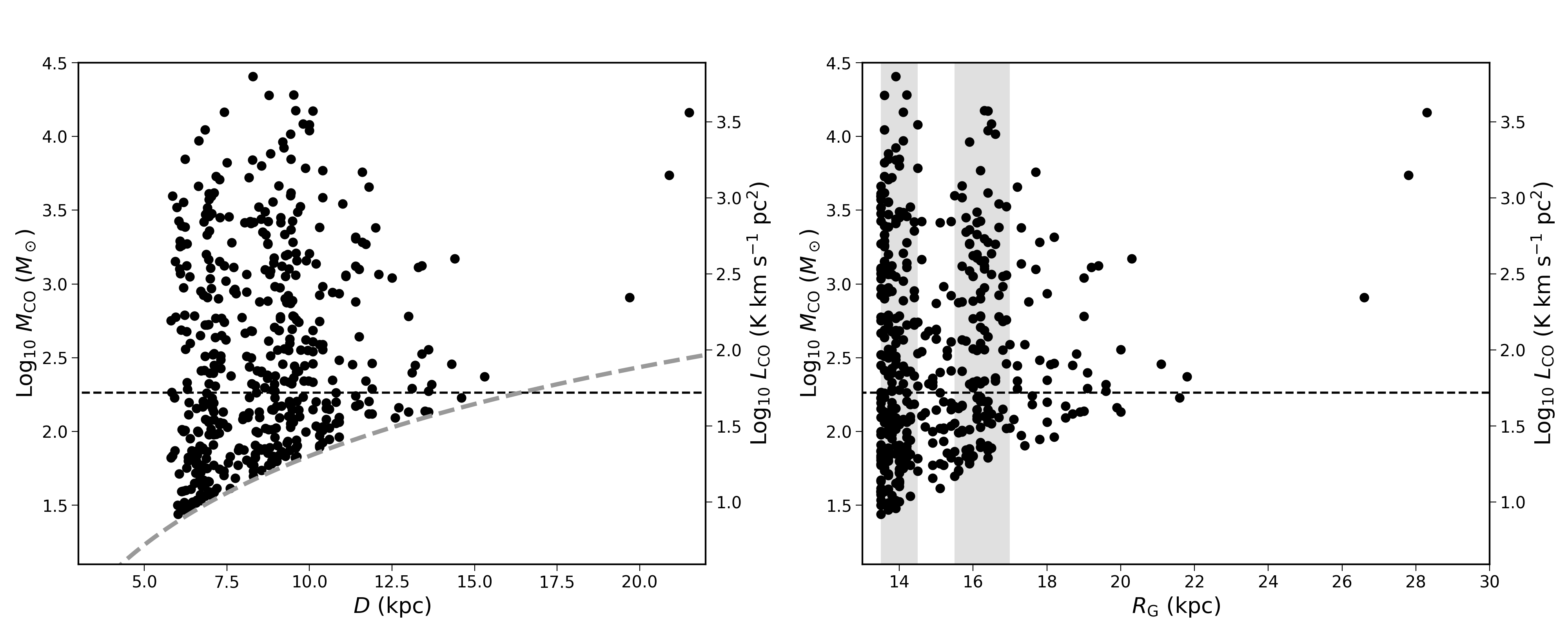}
\caption{
Left: Cloud mass (and CO luminosity) variation with kinematic distance ($D$) for all BKP clouds (466 clouds) in the outer Galaxy ($R_{\rm G}$ $\ge$ 13.5 kpc).
The gray dotted curve shows the minimum mass derived from the nominal completeness limit (2.64 K km s$^{-1}$) of the BKP catalog.
Right: Cloud mass (luminosity) variation with galactocentric radius ($R_{\rm G}$) for all BKP clouds (466 clouds) in the outer Galaxy ($R_{\rm G}$ $\ge$ 13.5 kpc).
The black dotted lines in the left and right panels show the minimum mass at $D$ = 16.4 kpc and mass threshold (183.6 $M_\odot$) for comparing the cloud properties at various distances up to $R_{\rm G}$ = 20.0 kpc.
The gray areas indicate the concentrated areas considered to be spiral arms.
Note that the two molecular clouds with kinematic distances of more than 20 kpc ($R_{\rm G}$ $>$ 27 kpc) are known to be actually located at a distance of 12 kpc ($R_{\rm G}$ = 19 kpc) from high-resolution optical spectra \citep[e.g.,][]{Smartt1996,Kobayashi2008}.
}
\label{Mass_completeness} 
\end{figure*}

The left panel of Figure \ref{Mass_completeness} shows the cloud mass (and luminosity) variation with $D$.
We estimated the masses of the BKP clouds ($M_{\rm CO}$) from the integrated CO line intensity ($I_{\rm CO}$).
We assumed a Galactic average mass-calibration ratio N(H$_2$)/$I_{\rm CO}$ of 2.0 $\times$ 10$^{20}$ cm$^{-2}$ (K km s$^{-1}$)$^{-1}$ \citep[e.g., ][]{Bolatto2013} 
and a correction for the helium abundance of 1.36 \citep[e.g.,][]{Kennicutt2012}.
The minimum mass (gray dotted curve in the left panel of Figure \ref{Mass_completeness}) was derived from the nominal completeness limit
(2.64 K km s$^{-1}$) of the BKP catalog.
At $D$ = 16.4 kpc, the minimum mass is 183.6 $M_\odot$ (black dotted lines in Figure \ref{Mass_completeness}).
The distance of 16.4 kpc corresponds to $R_{\rm G}$ = 20.0 kpc at $l$ = 102$^\circ$.49, which is the minimum $l$ value of the FCRAO survey
(in the $l$ range of the FCRAO survey, the distance corresponding to $R_{\rm G}$ = 20.0 kpc is largest at the minimum $l$ value).
Therefore, in the following sections, we only consider the clouds with a mass larger than
183.6 $M_\odot$ for comparison of molecular cloud properties at different $R_{\rm G}$ up to $R_{\rm G}$ = 20.0 kpc.
The right panel of Figure \ref{Mass_completeness} shows the cloud mass (and luminosity) variation with $R_{\rm G}$.
The massive molecular clouds larger than 10$^4$ $M_\odot$ are detected only at $R_{\rm G}$ $<$ 17 kpc.
This result is consistent with the surface density distribution of molecular gas in the outer Galaxy, wherein the densities decrease with increasing $R_{\rm G}$ \citep[e.g.,][]{Wolfire2003,Heyer2015,Nakanishi2016}.
Figure \ref{Mass_completeness} also shows that molecular clouds are concentrated around $R_{\rm G}$ $\sim$ 13.5--14.5 kpc and 15.5--17.0 kpc (the gray areas in the right panel of Figure \ref{Mass_completeness}).
Especially, massive clouds with $M_{\rm CO}$ of more than 10$^4$ $M_\odot$ are only located in these areas
\footnote{One molecular cloud with $M_{\rm CO}$ of more than 10$^4$ $M_\odot$ is located at $R_{\rm G}$ $\sim$ 28 kpc in Figure \ref{Mass_completeness}, but this cloud is known to be actually located at $R_{\rm G}$ = 19 kpc from high-resolution optical spectra \citep[e.g.,][]{Smartt1996, Kobayashi2008}. Therefore, the actual $M_{\rm CO}$ of this cloud is considered to be less than 10$^4$ $M_\odot$.}.
Therefore, these concentrated more densely populated areas and the less populated areas are considered to be the spiral arms and interarm areas, respectively. 

\subsection{WISE MIR data}\label{sec:2_2}
The candidate star-forming regions are \textit{WISE} MIR sources that meet our developed identification criterion (Section \ref{sec:1}).
The \textit{WISE} magnitudes and colors of the candidates are listed in the AllWISE Source Catalog\footnote{http://wise2.ipac.caltech.edu/docs/release/allwise/}\citep{https://doi.org/10.26131/irsa1} .
The AllWISE Source Catalog contains astrometry and photometry for 747,634,026 objects detected in the deep AllWISE Atlas Intensity Images \footnote{Detailed information is given at https://wise2.ipac.caltech.edu/docs/release/allwise/expsup/}. 

\subsubsection{Sensitivitiy of WISE data}\label{sec:2_21}
\begin{figure*}
\epsscale{1.0}
\plotone{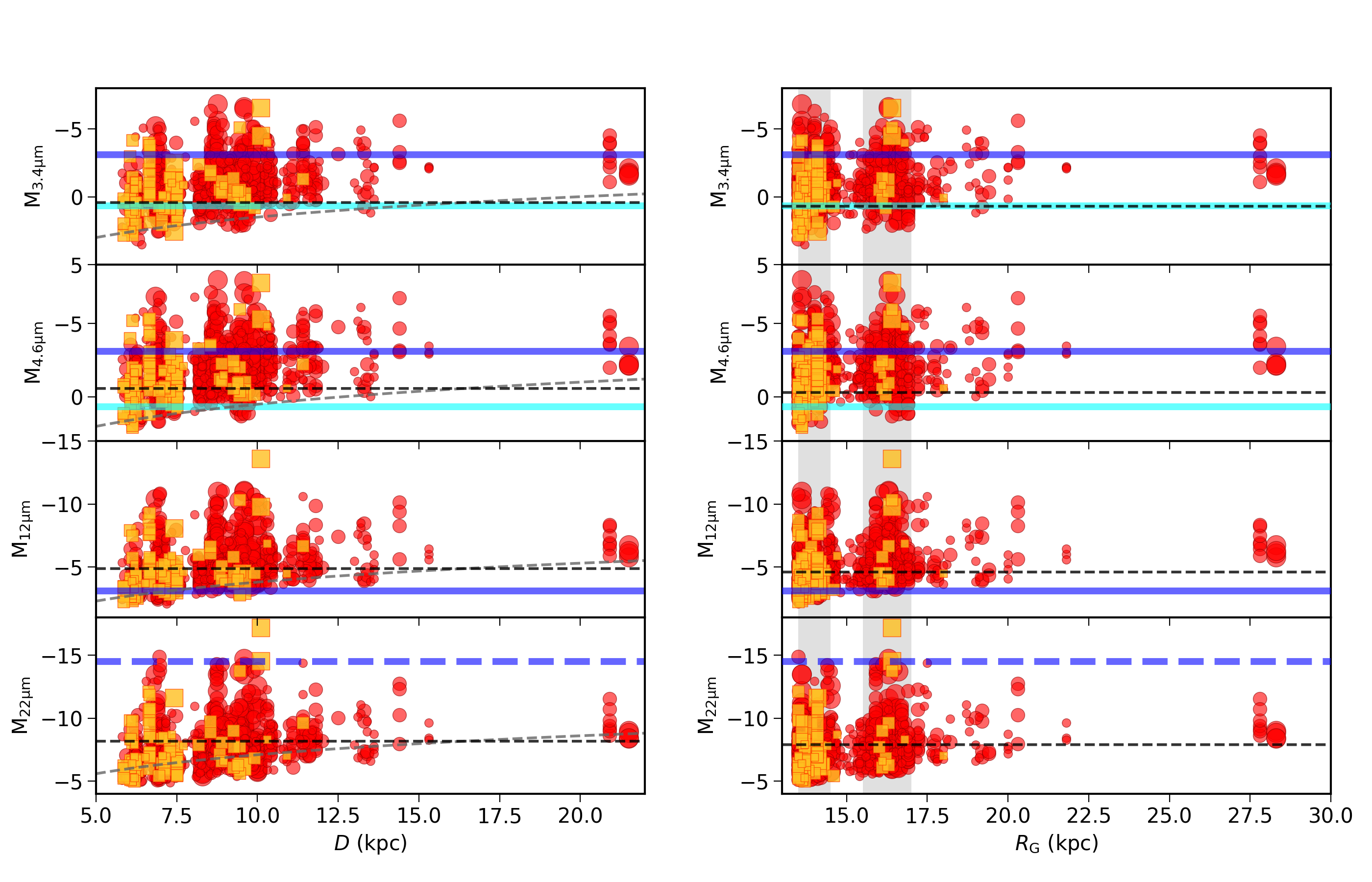}
\caption{
Left: Absolute magnitudes variation with kinematic distance ($D$) for the newly identified candidate star-forming regions in Paper I.
The grey dotted curves show the average detection limit for the minimum integration for eight frames
\citep[16.5, 15.5, 11.2, and 7.9 mag for 3.4, 4.6, 12, and 22 $\micron$, respectively; ][]{Wright2010}.
Right: Absolute magnitudes variation with galactocentric radius ($R_{\rm G}$) for the newly identified candidate star-forming regions in Paper I.
Red circles and yellow squares show candidates with a contamination rate of their parental clouds of $<$ 30 \% (reliable) and $\ge$ 30 \% (less reliable), respectively.
The size of these markers indicate the mass of their parental molecular clouds (small: $M_{\rm CO}$ $<$ 10$^3$ $M_\odot$,
middle: 10$^3$ $M_\odot$ $\le$ $M_{\rm CO}$ $<$ 10$^4$ $M_\odot$,
large: 10$^4$ $M_\odot$ $\le$ $M_{\rm CO}$).
The black dotted lines show the detection limit at $D$ = 16.4 kpc magnitude threshold for comparing the cloud and star formaion properties at various distances up to $R_{\rm G}$ = 20.0 kpc (M$_{\rm 3.4 \micron}$: 0.43 mag, M$_{\rm 4.6 \micron}$: -0.57 mag, M$_{\rm 12 \micron}$: -4.87 mag, M$_{\rm 22 \micron}$: -8.17 mag).
Cyan and blue lines show the absolute magnitude for the A0 and B0 stars in the main sequence \citep{Cox2000}.
Note that we used intrinsic $V$ $-$ $L$ colors (V band: 0.5555 $\micron$; L band: 3.547 $\micron$) and the absolute $V$-band magnitudes of A0 and B0 stars for calculating the apparent magnitudes of those stars for any of the four bands, since infrared colors, such as [3.4]$-$[22], are negligible for those early-type stars.
The blue dotted line shows the 22 $\mu$m absolute magnitude for the \ion{H}{2} regions ionized by B0 stars \citep{Anderson2014}.
Note that 11 star-forming regions with kinematic distances of more than 20 kpc ($R_{\rm G}$ $>$ 27 kpc) are known to be actually located at a distance of 12 kpc ($R_{\rm G}$ = 19 kpc) from high-resolution optical spectra \citep[e.g.,][]{Smartt1996,Kobayashi2008}.
The gray areas indicate the concentrated areas considered to be spiral arms.
} 
\label{WISE_completeness} 
\end{figure*}

The left panel of Figure \ref{WISE_completeness} shows the \textit{WISE} absolute magnitude variation with $D$
for the newly identified candidate star-forming regions in Paper I.
The detection limits (gray dotted curves in the left panel of Figure \ref{WISE_completeness}) were derived from 
the average detection limits for the minimum integration
for eight frames \citep[16.5, 15.5, 11.2, and 7.9 mag for 3.4, 4.6, 12, and 22 $\micron$, respectively; ][]{Wright2010}.
At $D$ = 16.4 kpc, which corresponds to $R_{\rm G}$ = 20.0 kpc (see details in Section \ref{sec:2_12}),
the detection limits for all four bands are
M$_{\rm 3.4 \micron}$ = 0.43, M$_{\rm 4.6 \micron}$ = -0.57,
M$_{\rm 12 \micron}$ = -4.87, and M$_{\rm 22 \micron}$ = -8.17.
At 3.4 $\micron$, this value roughly corresponds to the magnitude of the A0 star in the main sequence \citep[cyan line in Figure \ref{WISE_completeness};][]{Cox2000}.
The right panel of Figure \ref{WISE_completeness}, which shows the \textit{WISE} absolute magnitude variation with $R_{\rm G}$, indicates that the bright candidates with M$_{\rm 22 \micron}$ brighter than -14.5 mag are detected only at $R_{\rm G}$ $<$ 17 kpc. 
This value corresponds to the magnitude of the \ion{H}{2} region ionized by B0 stars \citep[blue dotted line in Figure \ref{WISE_completeness};][]{Anderson2014}.
Especially, such bright sources are only located in the spiral arms (Section \ref{sec:2_12})

\subsubsection{Re-identification of candidates}\label{sec:2_22}
In this paper, we redefine the candidate star-forming regions in order to compare the properties of molecular clouds and star-forming regions
at different $R_{\rm G}$ up to $R_{\rm G}$ = 20.0 kpc.
The candidates with absolute magnitudes at all four bands brighter than the detection limits are redefined.
Under this redefinition, we re-identified 282 candidate star-forming regions in 121 clouds out of 466 clouds.
Among all the 121 clouds, 108 clouds are considered to be reliable candidates
based on the adopted contamination threshold (see Section \ref{sec:1})
\footnote{We cannot re-calculate the contamination rate considering the thresholds with the same methods in the Paper I
because we cannot derive absolute magnitudes (distance) of \textit{WISE} sources in the field region.}.

\section{Properties of molecular clouds}\label{sec:3}
This section discusses the properties of BKP clouds with and without candidate star-forming regions in the outer Galaxy. 

\subsection{Mass distribution}\label{sec:3_1}
Figure \ref{Mass_spec} shows the number distribution of cloud mass (top panels) and 
the mass distributions (bottom panels) of
all BKP clouds in the outer Galaxy (left panels) and
BKP clouds with and without candidate star-forming regions in the outer Galaxy (right panels).
All mass distributions have been fitted to the power-law function:
$dN/dM$ $\propto$ $M^{- \alpha}$ ($N$: number, $M$: mass)
at $M_{\rm CO}$ $\ge$ 183.6 $M_\odot$.
The $\alpha$ value (corresponding to the slope) for all BKP clouds is 1.36 $\pm$ 0.08.
The $\alpha$ values for all BKP clouds with candidates and only BKP clouds with reliable candidates
are $\alpha$ = 1.04 $\pm$ 0.13 and $\alpha$ = 0.99 $\pm$ 0.13, respectively.
In contrast, the $\alpha$ value for BKP clouds without candidates is $\alpha$ = 1.57 $\pm$ 0.08.
The higher $\alpha$ value (corresponding to the steeper slope) for BKP clouds without associated candidates
indicates that stars tend to be born in higher mass clouds
($M_{\rm CO}$ $\ge$ 10$^3$ $M_\odot$).
This is also evident from the number distributions where the distributions with candidates clearly peak at higher $M_{\rm CO}$ (see top-right panel of Figure \ref{Mass_spec}).

In order to assess the possible impact on the above $\alpha$ values by the choice of the binning intervals for the mass $M_{\rm CO}$,
we compare the observed cumulative mass distribution with the simulated cumulative mass distributions (Figure \ref{cumMass_spec}).
The simulated clump samples are randomly generated within the same mass range  as that of the BKP clouds in the outer Galaxy ($M_{\rm CO} = $27.5--25,600.0 $M_\odot$).
Each simulated sample has 1,000,000 clumps, and the true mass functions for the samples are power laws with indices ranging from $\alpha$ = 0.0 to $\alpha$ = 2.0. 
The $\alpha$ value of the simulated cumulative mass function which best agrees with that of the observed all BKP clouds is 1.4.
For all BKP clouds with candidates, only BKP clouds with reliable candidates,
and BKP clouds without candidates,
the best agreed simulated $\alpha$ values are 1.0, 1.0, and 1.6, respectively.
These results are largely consistent with the results of power-law fittings for mass distributions,
suggesting that our choice of binning interval does not introduce any significant bias in the power-law slope.

\subsection{Relation between virial mass and mass derived from CO intensity}\label{sec:3_2}
The top panel of Figure \ref{Mvir_Mco} shows the relation between the virial mass ($M_{\rm vir}$) and the mass derived from CO intensity ($M_{\rm CO}$).
We estimated $M_{\rm vir}$ of the BKP clouds from $dv$ and $r$: $M_{\rm vir}$ = 210 $\times$ $r$ $dv ^2$,
with the assumption of uniform density clouds \citep[e.g.][]{Bertoldi1992,Heyer2001}.
In the BKP catalog, the Gaussian fits were attempted for all clouds defined over at least three spectroscopic channels.
Among all the 466 clouds, the fitting failed to converge or was not attempted for 277 clouds
due to their small number of spectroscopic channels.
Therefore, we could estimate $M_{\rm vir}$ of only 189 clouds out of 466 clouds.
(see Section \ref{sec:2_11}).
Also, we note that $r$ is estimated from $l_{\rm major}$ and $l_{\rm minor}$ which directly translates into the virial mass estimate.
We plot the upper limit of the $M_{\rm vir}$ for the other clouds using the velocity resolution ($dv$ = 0.98 km s$^{-1}$). 
The BKP clouds with $M_{\rm CO}$ $\ge$ 10$^{3}$ $M_\odot$ roughly follow the line of $M_{\rm vir}$ = $M_{\rm CO}$,
while the clouds with $M_{\rm CO}$ $<$ 10$^{3}$ $M_\odot$ are concentrated in the region of $M_{\rm vir}$ $>$ $M_{\rm CO}$.
The former group is mainly composed of clouds with candidate star-forming regions, and the latter group is mainly composed of clouds without candidates.
These results indicate that almost all clouds with candidates are bound by self-gravity, while many clouds without candidates are not bound.
We checked the high-mass clouds ($\ge$ 10$^3$ $M_\odot$) without candidates to find that many of them have faint candidates with absolute magnitudes smaller than the
detection limits of \textit{WISE} data
(those are removed from the candidates in this paper, see Section \ref{sec:2_22}).
Thus, we suggest that the other clouds bound by self-gravity without candidates also have faint candidates, which are fainter than the detection limit of the the \textit{WISE} data.
We note that only less than 50 \% of clouds with $M_{\rm CO}$ $<$ 10$^{3}$ $M_\odot$ have linewidths larger than the velocity resolution (bottom panel of Figure \ref{Mvir_Mco}).
Higher velocity resolution data are crucial to investigate the relation between $M_{\rm vir}$ and $M_{\rm CO}$ for low-mass clouds.

In this paper, we derived $M_{\rm CO}$ assuming that the calibration rate N(H$_2$)/$I_{\rm CO}$ in the outer Galaxy ($R_{\rm G}$ = 13.5 -- 20 kpc) is similar to that in the solar neighborhood.
If the mass calibration rate in the outer Galaxy is different from that in the solar neighborhood, 
BKP clouds are considered to not follow the line of $M_{\rm vir}$ = $M_{\rm CO}$ and their $M_{\rm CO}$ is mostly scaling (i.e., shifting along the horizontal axis in Figure \ref{Mvir_Mco}) depending on the actual mass calibration rate.
Therefore, these results also suggest that the mass calibration rate in the outer Galaxy 
is similar to that in the solar neighborhood,
although the metallicity in the outer Galaxy is less than about one third of that in the solar neighborhood \citep[e.g.,][]{Smartt1997,Fernandez2017}.
To detect clear differences in the mass calibration rate from the solar neighborhood, we may need to detect
enough molecular clouds
at $R_{\rm G}$ $\ge$ 18 kpc, where the metallicity is less than about one fifth of that in the solar neighborhood \citep[e.g.][]{Smartt1997,Bolatto2013,Fernandez2017}.

\subsection{Relation between size and linewidth}\label{sec:3_3}
Figure \ref{Larson} shows the size-linewidth ($r$-$dv$) relation of BKP clouds in the outer Galaxy.
Note that this figure shows only 189 clouds out of 466 clouds because the linewidths of the other 277 clouds are not derived successfully in the BKP catalog (see Section \ref{sec:3_2}).
For the other clouds, we plot the velocity resolution ($dv$ = 0.98 km s$^{-1}$) as the upper limit of the linewidth.
The distribution range of size and linewidth for clouds with candidates
(2 $\lesssim$ $r$ $\lesssim$ 20 pc, 0.6 $\lesssim$ $dv$ $\lesssim$ 4 km s$^{-1}$)
is similar to those for clouds without candidates.
We compare this result with the results of previous studies for molecular clouds in the
outer Galaxy \citep[$R_{\rm G}$ $\gtrsim$ 16 kpc; solid gray line in Figure \ref{Larson};][]{Brand1995},
inner Galaxy \citep[$R_{\rm G}$ $<$ 16 kpc; black dotted line in Figure \ref{Larson};][]{Brand1995},
and dwarf galaxies \citep[blue dotted line in Figure \ref{Larson};][]{Rubio2015}.
As a result, we confirmed that the BKP clouds in the outer Galaxy roughly follow
the least-square fits of previous studies
for molecular clouds in the outer Galaxy and dwarf galaxies.

\subsection{Relation between size and mass}\label{sec:3_4}
Figure \ref{Mco_r} shows the size-mass ($r$-$M_{\rm CO}$) relation of all 466 BKP clouds in the outer Galaxy.
The Figure also shows the histograms, which indicate the number distribution of size (Figure \ref{Mco_r}, b) and mass (Figure \ref{Mco_r}, c) for BKP clouds with and without candidates.
In the size number distribution, the peak of the histogram for BKP clouds with and without candidates is about
3.2 pc and 1.6 pc, respectively.
While in the mass number distribution, the peak of the histogram for BKP clouds with and without candidates is about
1700 $M_\odot$ and 130 $M_\odot$ (less than the mass threshold), respectively.
It means that the size of BKP clouds with candidates is about twice as large as that without candidates,
while the mass of BKP clouds with candidates is more than ten times as large as that without candidates.
This result suggests that the BKP clouds with candidates have a larger column density 
than twice the BKP clouds without candidates.

\section{Properties of star-forming regions}\label{sec:4}
In this section, we discuss the properties of 282 candidate star-forming regions in 121 BKP clouds.

\subsection{Relation between color and cloud mass}\label{sec:4_1}
Figure \ref{Color-mass} shows the relation between \textit{WISE} colors of candidate star-forming regions: [3.4] - [4.6], [4.6] - [12], [4.6] - [22],
and mass of their parental BKP clouds.
The right panel of Figure \ref{Color-mass} presents the \textit{WISE} colors of 
282 individual candidates,
while the left panel of Figure \ref{Color-mass} displays the \textit{WISE} colors of
121 integrated (total) candidates in each parental cloud.
The candidate star-forming regions are broadly distributed at 0.5 $\le$ [3.4] - [4.6] $\le$ 3, 2$\le$ [4.6] - [12] $\le$ 6, and 3 $\le$ [4.6] - [22] $\le$ 10.
The correlation coefficients (CCs)\footnote{
In this paper, we used the Peason's Correlation Coefficient.
} between \textit{WISE} colors and the cloud mass of all plots in Figure \ref{Color-mass}
range from -0.050 to 0.116 (see Figure \ref{Color-mass}).
For massive clouds, reddening of [4.6] - [12] and slight blueing of [3.4] - [4.6] was expected due to polycyclic aromatic hydrocarbon (PAH) emission,
which is known to be strong at 12$\micron$ and 3.4 $\micron$ for massive star-forming regions with OB stars \citep{Wright2010}.
However, these CCs indicate that there is no correlation between \textit{WISE} colors and cloud mass. 

\subsection{Relation between luminosity and cloud mass}\label{sec:4_2}
Figure \ref{Mag-mass} shows the relation between monochromatic luminosities of candidate star-forming regions and their parental BKP cloud mass.
The right panel of Figure \ref{Mag-mass} shows the monochromatic luminosities of individual candidates. 
In contrast, the left panel of Figure \ref{Mag-mass} shows the integrated (total) monochromatic luminosities of candidates in each parental cloud.
The monochromatic luminosities were calculated from \textit{WISE} magnitudes in the AllWISE catalog
with the Zero Magnitude Flux Density of WISE data \citep[$F_{3.4 \micron}$ = 309.540, $F_{4.6 \micron}$ = 171.787, $F_{12 \micron}$ = 31.674, and $F_{22 \micron}$ = 8.363 Jy:][]{Jarrett2011}
and kinematic distance ($D$) of their parental clouds.
Figure \ref{Mag-mass} indicates that brighter candidates:
$L_{\rm 3.4 \micron}$ $\ge$ 10$^{15}$ W Hz$^{-1}$,
$L_{\rm 4.6 \micron}$ $\ge$ 10$^{15}$ W Hz$^{-1}$,
$L_{\rm 12 \micron}$ $\ge$ 10$^{16}$ W Hz$^{-1}$,
and $L_{\rm 22 \micron}$ $\ge$ 10$^{17}$ W Hz$^{-1}$,
are only associated with higher mass clouds larger than 10$^3$ $M_\odot$. 
The threshold of $L_{\rm 22 \micron}$ roughly corresponds to the luminosity of an \ion{H}{2} region ionized by B0 star \citep[blue dotted lines in Figure \ref{Mag-mass};][]{Anderson2014}.

\section{Star-formation activities}\label{sec:5}
In this section, we discuss the variation of star-formation activities at different $R_{\rm G}$.

\subsection{Spiral arm versus inter-spiral arm areas}\label{sec:5_new}
At $R_{\rm G}$ = 20.0 kpc, \ion{H}{1} gas surface density, H$_2$ gas surface density, and metallicity are less than half of those at $R_{\rm G}$ = 13.5 kpc
\citep[e.g.,][]{Wolfire2003,Heyer2015,Fernandez2017}.
Furthermore, we reported that the $R_{\rm G}$ $\sim$ 13.5--14.5 (arm-1)
and 15.5--17.0 kpc (arm-2) areas are considered to be the spiral arms (Section \ref{sec:2_12}; gray areas in Figures \ref{Mass_completeness} and \ref{WISE_completeness}).
These spiral arm distributions are shown in Figure \ref{XY}.
In order to discuss these structures quantitatively, we investigate the number distribution of cloud mass in these areas (Figure \ref{hist-arm}).
The peaks above the mass threshold (183.6 $M_\odot$) are around 10$^{3.5}$ $M_\odot$ and 10$^{3}$ $M_\odot$ for arm-1 and arm-2.
There is no peak for clouds in the other area (interarm area) above the mass threshold.
The number is decreasing with cloud mass in the interarm area.
We perform a Kolmogorov-Smirnov (KS) test to compare the number distribution of arm-1, arm-2, and interarm area.
The KS test returns a probability (p-value) that the two samples came from identical populations.
We adopt a p-value of 0.05 for the null hypothesis that two distributions are identical.
If the p-value is smaller than 0.05, we reject the null hypothesis.
The derived p-values for arm-1 versus interarm, arm-2 versus interarm, and arm-1 versus arm-2 to be identical, are 0.0048, 0.0032, and 0.62, respectively.
Hence, the mass distribution in the two spiral arms (arm-1 and arm-2) and the interarm areas are different,
while the distributions in the two spiral arms are identical.

\subsection{Relation between WISE color and $R_{\rm G}$}\label{sec:5_1}
Figure \ref{Color-rg} shows relations between \textit{WISE} colors ([3.4] - [4.6], [4.6] - [12], [4.6] - [22])
of candidate star-forming regions and $R_{\rm G}$.
Candidates are distributed in the color range of 0.5 $\le$ [3.4] - [4.6] $\le$ 3, 2 $\le$ [4.6] - [12] $\le$ 6, and 3 $\le$ [4.6] - [22] $\le$ 10.
It is found that the very red sources with  [3.4] - [4.6] $>$ 1.5, [4.6] - [12] $>$ 4.0, or [4.6] - [22] $>$ 7.0
are mostly present at $R_{\rm G}$ $<$ 18 kpc, especially in the spiral arms, 
and almost all sources at $R_{\rm G}$ $\ge$ 18 kpc are bluer than those colors.
Note that this “blueing” toward larger $R_{\rm G}$ could be due to the stochastic effect with the small number of sources at $R_{\rm G}$ $\ge$ 18 kpc (bottom panels of Figure \ref{Color-rg}), and further study with more surveys is desirable. 
However, if the blueing trend is a real feature,
it could be interpreted as a result of the absence of massive star-forming regions at $R_{\rm G}$ $\ge$ 18 kpc
considering the redding of [4.6] - [12] due to the PAH emission.
This is consistent with other results in the previous sections (see Sections \ref{sec:2_12} and \ref{sec:2_21}).
Figures \ref{Mass_completeness} and \ref{WISE_completeness} show that
massive clouds ($\ge$ 10$^4$ $M_\odot$) and luminous star-forming regions (22 $\micron$ absolute magnitude is brighter than
that of the \ion{H}{2} regions ionized by B0 stars)
are absent at $R_{\rm G}$ $\ge$ 18 kpc.
As such, the far outer region ($R_{\rm G}$ $\ge$ 18 kpc) appears to be devoid of massive star-formation, which is consistent with the absence of H$\alpha$ emission, which traces massive star-forming regions, in extra-galactic XUV disk \citep[e.g.,][]{Thilker2005}.
It also could be interpreted as a result of the
short lifetime of the circumstellar disk in the lower metallicity environment \citep[e.g.,][]{Yasui2010,Guarcello2021}.
This interpretation is based on the consideration that the redding of [3.4] - [4.6] and [4.6] - [12]
is due to excess emission from circumstellar disk/envelope material in young stellar objects \citep[e.g.,][]{Koenig2014}.

\clearpage
\subsection{Star formation efficiency}\label{sec:5_2}
Next, we discuss the SFE.
While star formation consists of two basic processes:1) conversion from \ion{H}{1} gas to H$_2$ gas and 2) conversion from H$_2$ gas to stars, our study focuses on the latter process as a first step.
To investigate the SFE, which represents the conversion of H$_2$ gas mass to stellar mass per molecular cloud,
we use the following two parameters constructed only from the \textit{WISE} MIR and FCRAO CO data:
1) the fraction of BKP clouds with candidates ($N_{\rm SF}$/$N_{\rm all}$),
and 2) the monochromatic MIR luminosities of the candidates per parental BKP cloud mass ($L_{\rm MIR}$/$M_{\rm CO}$).
Although these parameters are not entirely conclusive, they can provide a useful measure of SFE per molecular cloud. 

\subsubsection{$N_{\rm SF}$/$N_{\rm all}$}\label{sec:5_21}
The $N_{\rm SF}$/$N_{\rm all}$ is the simplest parameter of SFE averaged over all kinds of parameters (e.g., mass, age).
The large statistical number of candidate star-forming regions in our data set enables this study for the first time. 
The lower plot of Figure \ref{Total_rate_rg} shows the relation between the number of clouds with and without candidates and $R_{\rm G}$,
while the upper plot of Figure \ref{Total_rate_rg} shows the $R_{\rm G}$ variation of $N_{\rm SF}$/$N_{\rm all}$.
In order to compare the $N_{\rm SF}$/$N_{\rm all}$ at different $R_{\rm G}$, only clouds larger than 183.6 $M_\odot$ are plotted in Figure \ref{Total_rate_rg} (see Section \ref{sec:2_12}).
The least-squares fittings are performed at $R_{\rm G}$ $\le$ 20.0 kpc.
The fitting results for all clouds with candidates and only clouds with reliable candidates are
$N_{\rm SF}$/$N_{\rm all}$ = 5.8($\pm$2.5)$R_{\rm G}$ - 52.3($\pm$36.5)
and $N_{\rm SF}$/$N_{\rm all}$ = 6.2($\pm$2.1)$R_{\rm G}$ - 64.0($\pm$31.0), respectively.
These results suggest that $N_{\rm SF}$/$N_{\rm all}$ slightly increases with increasing $R_{\rm G}$, especially for the $R_{\rm G}$ range of 13.5 -- 18.0 kpc.
On a speculative note, this could hint at the presence of CO-dark clouds, which are difficult to detect by CO emission lines.
They are known to increase with decreasing metallicity, in other words, increasing $R_{\rm G}$ \citep[e.g.,][]{Wolfire2010}.
In this paper, $N_{\rm all}$ and $N_{\rm SF}$ mean the number of molecular clouds detected in CO emission.
Therefore, if we could detect molecular clouds in H$_2$ emission, the fraction of molecular clouds with candidates, defined as $N_{\rm SF-H_2}$/$N_{\rm all-H_2}$, is possibly constant (or decreases with increasing $R_{\rm G}$).
From these results we conclude that $N_{\rm SF}$/$N_{\rm all}$ does not decrease with increasing $R_{\rm G}$.
Since the lower plot in Figure \ref{Total_rate_rg} shows a significant measured difference in $N_{\rm all}$ with
$R_{\rm G}$ (as discussed in Sections \ref{sec:2_12} and \ref{sec:5_new}),
but the upper plot shows no significant variation in the ratio with $R_{\rm G}$,
it implies that the absolute number of BKP (lower plot) does not bias the $N_{\rm SF}$/$N_{\rm all}$.

Figure \ref{Mass_rate_rg} shows the same plot as Figure \ref{Total_rate_rg} but in three cloud mass ranges: 
$M_{\rm CO}$ $<$ 10$^3$ $M_\odot$ (left),
10$^3$ $M_\odot$ $\le$ $M_{\rm CO}$ $<$ 10$^4$ $M_\odot$ (middle), and 10$^4$ $M_\odot$ $\le$ $M_{\rm CO}$ (right).
In all three cases, we find that $N_{\rm SF}$/$N_{\rm all}$ does not decrease
with increasing $R_{\rm G}$.
Furthermore, the ratio increases with cloud mass from 20 -- 60 \% ($M_{\rm CO}$ $<$ 10$^3$ $M_\odot$)
to 40 -- 100 \% ($M_{\rm CO}$ $\ge$ 10$^3$ $M_\odot$).
This result is consistent with the finding that stars tend to be born in higher-mass clouds, as discussed in Section \ref{sec:3_1}.

\subsubsection{$L_{\rm MIR}$/$M_{\rm CO}$}\label{sec:5_22}
In past studies, SFE was measured by
the ratio of MIR-FIR luminosity ($\lambda$ = 12 -- 100 $\micron$ from $IRAS$ data) to molecular mass \citep[e.g.,][]{Snell2002}.
In this paper, we measure SFE with $L_{\rm MIR}$/$M_{\rm CO}$ by making use of the \textit{WISE} and FCRAO data.
Although the bolometric luminosity should be ultimately used for estimating the integrated luminosity, we use the monochromatic luminosity.
We note that emission in all four bands is known to be affected not only by radiation from dust warmed by star-forming activity but also by contamination from a variety of other dust sources:
3.4 and 12 $\micron$ include prominent PAH emission features, and the 4.6 $\micron$ measures the continuum emission from very small grains and the 22 $\micron$ represents both stochastic emission from small grains and the Wien tail of thermal emission from large grains \citep{Wright2010}.
However, particularly at shorter wavelengths (3 -- 5 $\micron$), the dust emission is less contaminated than at longer wavelengths
\citep[$\lambda$ $\ge$ 10 $\micron$; e.g., ][]{Popescu2011},
making it a useful indicator of the luminosity of star clusters and stellar aggregates.
Furthermore, these wavelengths are also valuable because they are much less affected by extinction than the shorter NIR.

Figure \ref{Lumimass_rg} shows the $R_{\rm G}$ variation of $L_{\rm MIR}$/$M_{\rm CO}$ for individual candidates (right panel)
and integrated candidates in each parental BKP cloud (left panel).
The values of $L_{\rm MIR}$/$M_{\rm CO}$ spread widely
over 3 to 4 orders of magnitude:
10$^{8}$ $\le$ $L_{\rm 3.4 \mu m}$/$M_{\rm CO}$ $\le$ 10$^{12}$ W Hz$^{-1}$ $M_\odot^{-1}$,
10$^{8}$ $\le$ $L_{\rm 4.6 \mu m}$/$M_{\rm CO}$ $\le$ 10$^{12}$ W Hz$^{-1}$ $M_\odot^{-1}$,
10$^{10}$ $\le$ $L_{\rm 12 \mu m}$/$M_{\rm CO}$ $\le$ 10$^{13}$ W Hz$^{-1}$ $M_\odot^{-1}$,
and 10$^{10}$ $\le$ $L_{\rm 22 \mu m}$/$M_{\rm CO}$ $\le$ 10$^{14}$ W Hz$^{-1}$ $M_\odot^{-1}$.
The CCs between $L_{\rm MIR}$/$M_{\rm CO}$ and $R_{\rm G}$ for all four bands range from 0.063 to 0.162 (see Figure \ref{Lumimass_rg}).
These small values indicate that there are no obvious trend with $R_{\rm G}$.
Furthermore, the panels in Figure \ref{Lumimass_rg} illustrate that the $L_{\rm MIR}$/$M_{\rm CO}$ distributions
are similar at any $R_{\rm G}$, partly represented by the constancy of maximum, mean, and median values of
$L_{\rm MIR}$/$M_{\rm CO}$.

\subsubsection{Environmental dependence of star formation efficiency}\label{sec:5_23}
We confirmed that the two SFE parameters,
$N_{\rm SF}$/$N_{\rm all}$ and $L_{\rm MIR}$/$M_{\rm CO}$,
do not decrease with increasing $R_{\rm G}$ at $R_{\rm G}$ of 13.5 kpc to 20.0 kpc.
Considering the possible effect from the presence of CO-dark clouds (Section \ref{sec:5_21}),
we find that these parameters do not show a clear evidence of change 
from $R_{\rm G}$ = 13.5 to 20.0 kpc.
Interestingly, no variation is found between spiral-arm regions
(arm-1 and arm-2, the gray areas in the right panel of Figure \ref{Mass_completeness}) and interarm regions.
This result suggests that the SFE per molecular cloud does not depend on the environmental parameters, such as metallicity and gas surface densities, which vary considerably with the $R_{\rm G}$.
Also, this result is consistent with the previous study by \citet{Snell2002}.
They investigated the ratio of MIR -- FIR luminosity to molecular cloud mass in the outer Galaxy using 23 \textit{IRAS} sources and 246 molecular clouds ($R_{\rm G}$ $\ge$ 13.5 kpc) detected from the FCRAO survey \citep{Heyer1998} and found that the ensemble value is similar to that in the W3/W4/W5 cloud complex.

Previous studies reported that SFE/yr converting from entire gas mass to stellar mas
(derived from $\Sigma_{\rm SFR}$/$\Sigma _{\rm gas}$)
decreases with increasing $R_{\rm G}$ at $R_{\rm G}$ $\gtrsim$ 13.5 kpc ($\Sigma _{\rm gas}$ $\lesssim$ 10 $M_\odot$pc$^{-2}$) in the outer Galaxy \citep[e.g.,][]{Kennicutt2012}.
Based on the assumption that SFE is proportional to the SFE/yr, our result suggests that the decrease of the SFE/yr is due to the decrease in conversion efficiency of \ion{H}{1} gas mass to H$_2$ gas mass.
In the inner Galaxy ($R_{\rm G}$ $<$ 13.5 kpc, $\Sigma _{\rm gas}$ $\gtrsim$ 10 $M_\odot$pc$^{-2}$),
SFE/yr converting from H$_2$ gas mass to stellar mass (derived from $\Sigma_{\rm SFR}$/$\Sigma _{\rm H_2}$)
is reported to be constant although $\Sigma _{\rm H_2}$ decreases with increasing $R_{\rm G}$.
Therefore, our result suggests that this trend reported in the inner Galaxy also holds for the outer Galaxy. 

For nearby spiral galaxies, recent studies of the PHANGS (Physics at High Angular resolution in Nearby GalaxieS) project with ALMA reported that
SFE/yr converting from H$_2$ gas mass to stellar mass is roughly constant for
0.5 $\lesssim$ $\Sigma_{\rm H_2}$ $\lesssim$ 100 $M_\odot$ pc$^{-2}$
\citep{Querejeta2021}.
This result is consistent with the results of the previous studies for external galaxies
\citep[3 $\lesssim$ $\Sigma_{\rm H_2}$ $\lesssim$ 100 $M_\odot$ pc$^{-2}$; e.g.,][]{Bigiel2010}.
Our result suggests that this trend also holds for 0.1 $\lesssim$ $\Sigma_{\rm H_2}$ $\lesssim$ 1 $M_\odot$ pc$^{-2}$ \citep[e.g.][]{Heyer2015,Nakanishi2016}.

We note that our main conclusion in this section (SFE converting from H$_2$ gas mass to stellar mass does not change through $R_{\rm G}$ = 13.5 -- 20.0 kpc)
is not affected by the threshold values of the FCRAO (Section \ref{sec:2_12}) and \textit{WISE} data (Section \ref{sec:2_22}). 
Nevertheless, the actual values of SFE parameters can be different and are mostly scaled depending on the threshold values (i.e., number of sources selected).
This is further detailed in Appendix \ref{sec:a_1} where we present the two SFE-parameters derived from several thresholds.

\section{Summary} \label{sec:6}
\begin{figure*}
\epsscale{1.0}
\plotone{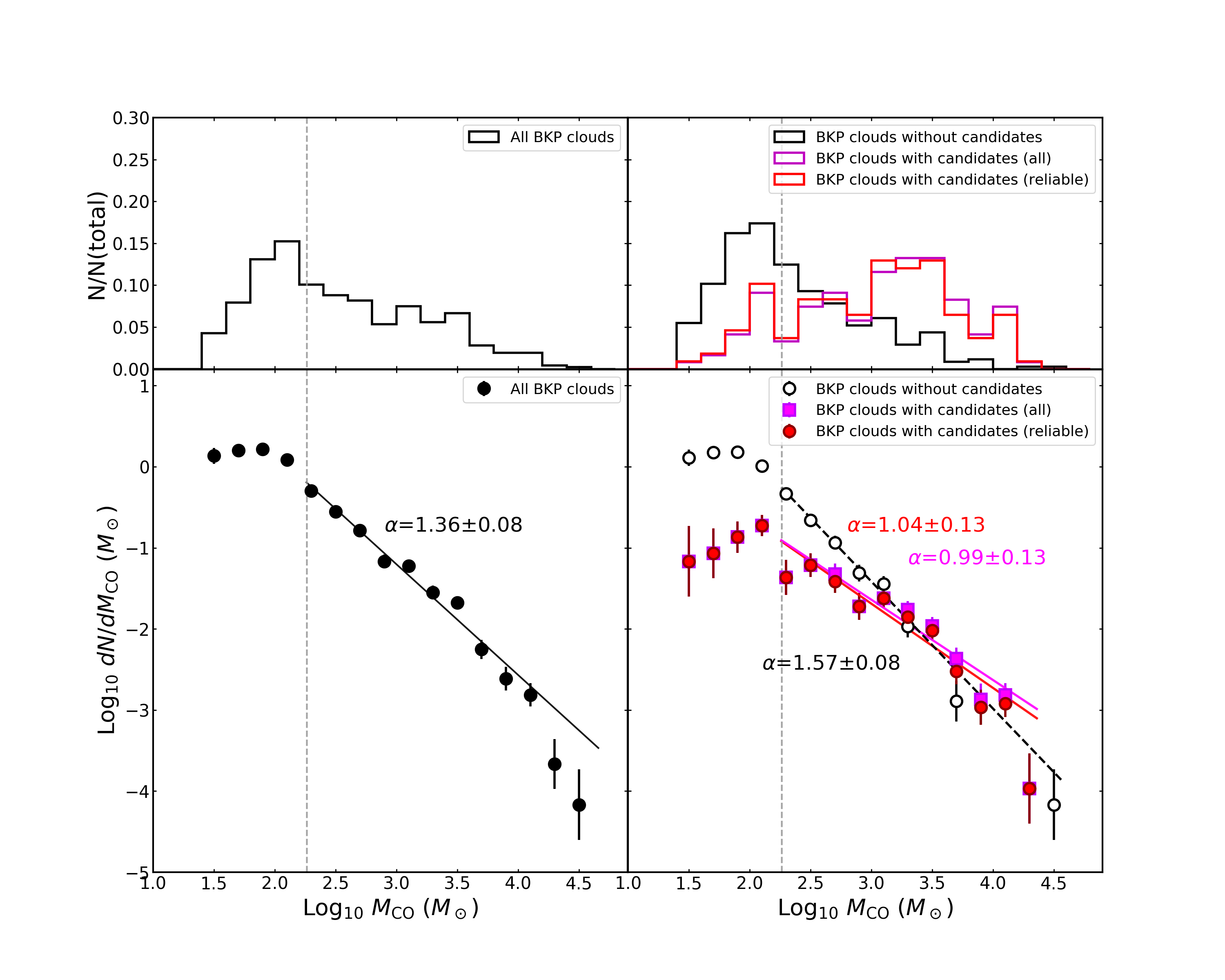}
\caption{
Top-left:
Normalized number distribution of cloud mass for all 466 BKP clouds in the outer Galaxy.
Top-right:
Normalized number distributions of cloud mass for clouds without candidates (black),
clouds with all associated candidates (magenta),
and clouds with only reliable associated candidates (red).
Bottom-left:
Cloud mass distribution of all 466 BKP clouds in the outer Galaxy.
The black line shows the result of the least-squares fitting with errors for a cloud mass of more than 183.6 $M_\odot$:
log$_{10}$ $dN/dM_{\rm CO}$ = -1.36($\pm$0.08)$\times$log$_{10} M_{\rm CO}$
+ 2.89($\pm$0.23).
The vertical gray-dashed line indicates the mass threshold (183.6 $M_\odot$). 
The binning is in logscale, into 25 intervals covering the mass range 10$^0$ to 10$^{4.8}$ $M_\odot$.
Bottom-right:
Cloud mass distribution of clouds with and without candidate star-forming regions.
Magenta squares and black open circles show clouds with and without candidates, respectively.
Red circles show only clouds with reliable candidates.
The magenta, red, and black-dotted lines show
the result of the least-squares fitting with errors for 
a cloud mass of more than 183.6  $M_\odot$:
log$_{10}$ $dN/dM_{\rm CO}$ = -1.04($\pm$0.13)$\times$log$_{10} M_{\rm CO}$
+ 1.43($\pm$0.42) (magenta: all clouds with candidates),
log$_{10}$ $dN/dM_{\rm CO}$ = -0.99($\pm$0.13)$\times$log$_{10} M_{\rm CO}$
+ 1.33($\pm$0.42) (red: only clouds with reliable candidates),
and
log$_{10}$ $dN/dM_{\rm CO}$ = -1.57($\pm$0.08)$\times$log$_{10} M_{\rm CO}$
+ 3.29($\pm$0.22) (black: clouds without candidates).
The error bars show Poisson counting uncertainties.
}  
\label{Mass_spec} 
\end{figure*}

\begin{figure*}
\epsscale{1.0}
\plotone{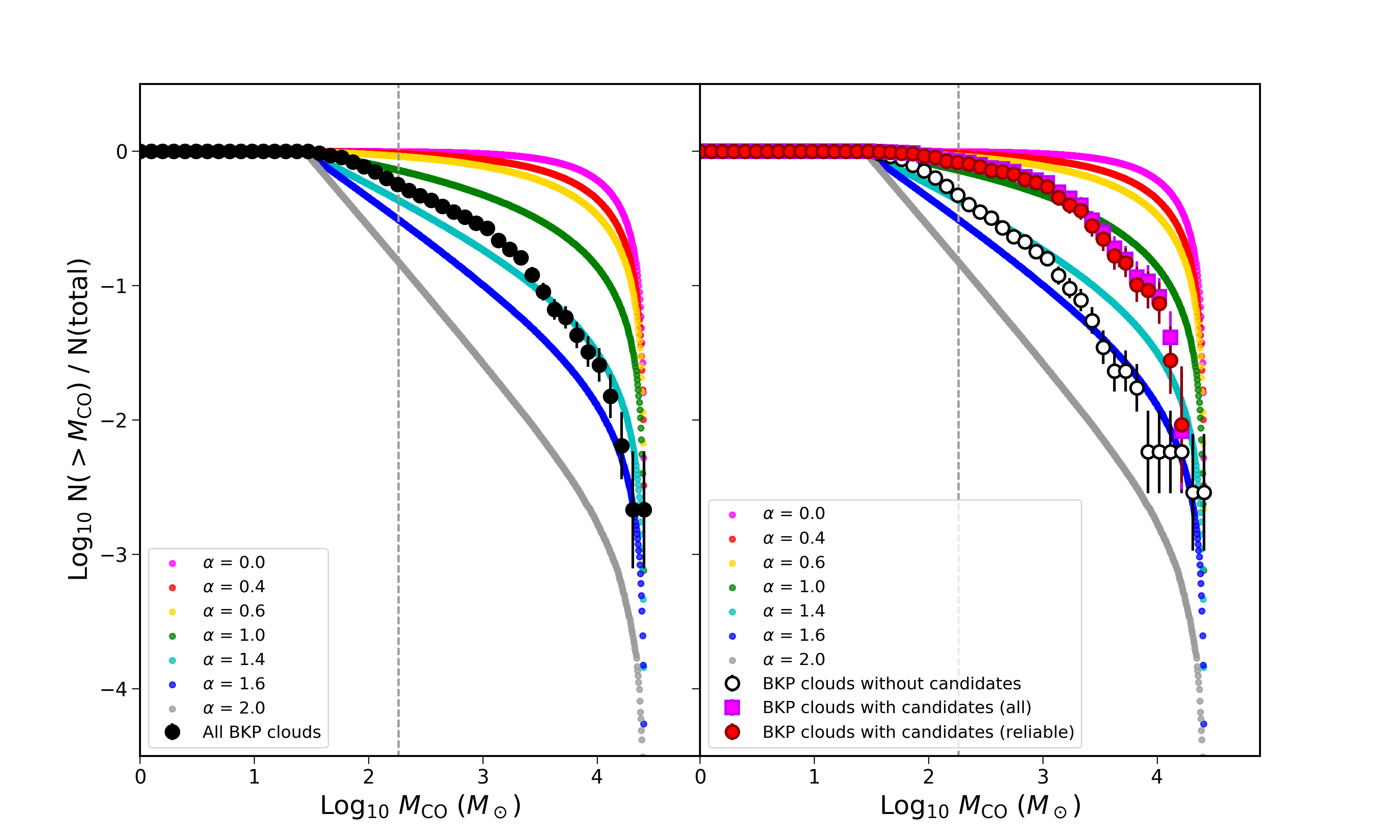}
\caption{
Left: Cumulative cloud mass distribution of all 466 BKP clouds in the outer Galaxy.
Right: Cumulative cloud mass distribution of clouds with and without candidate star-forming regions.
Magenta squares and black open circles show clouds with and without candidates, respectively.
Red circles show only clouds with reliable candidates.
The error bars show Poisson counting uncertainties.
The vertical gray-dashed line indicates the mass threshold (183.6 $M_\odot$). 
Magenta, red, yellow, green, cyan, blue, and gray points show the simulated results
(magenta: $\alpha$ = 0.0, red: $\alpha$ = 0.4, yellow: $\alpha$ = 0.6, green: $\alpha$ = 1.0, cyan: $\alpha$ = 1.4, blue: $\alpha$ = 1.6, gray: $\alpha$ = 2.0).
}  
\label{cumMass_spec} 
\end{figure*}

\begin{figure*}
\epsscale{1.0}
\plotone{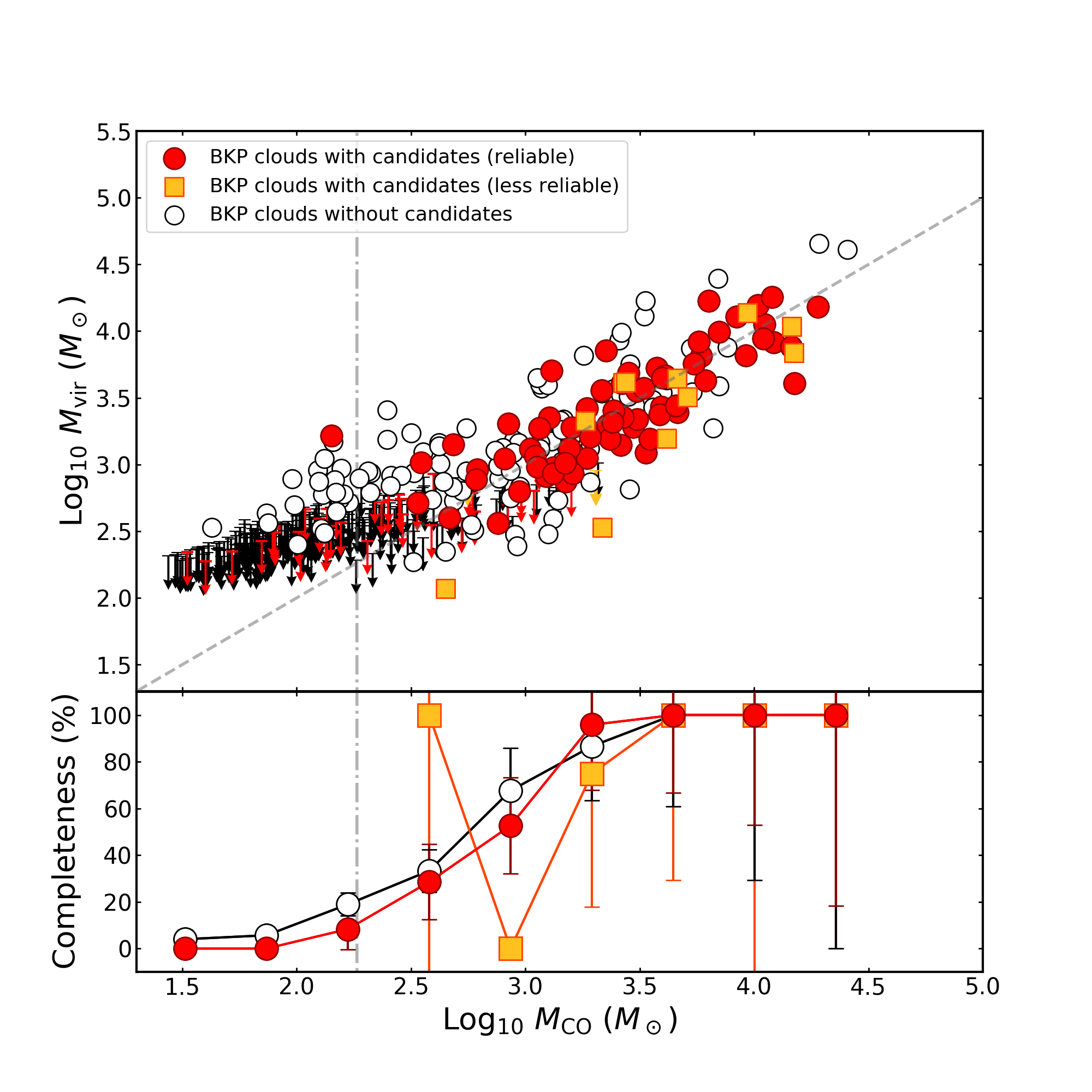}
\caption{
Top: Relation between virial mass ($M_{\rm vir}$) and mass derived from CO intensity ($M_{\rm CO}$).
Red circles and yellow squares show clouds with associated candidate star-forming regions with a contamination rate of their parental clouds
of $<$ 30 \% (reliable) and $\ge$ 30 \% (less reliable), respectively.
Black open circles show clouds without associated star-forming regions.
The gray-dotted line shows the $M_{\rm vir}$ = $M_{\rm CO}$
relation.
The vertical gray-dotdashed line indicates the mass threshold (183.6 $M_\odot$). 
The arrows indicate the upper limit (using $dv$ = 0.98 km s$^{-1}$) for 277 clouds
for which the linewidths are not derived in BKP catalog.
Bottom: Relation between completeness of $M_{\rm vir}$ (number of clouds whose linewidth is derived / total number of clouds) and $M_{\rm CO}$.
The error bars represent Poisson errors (1 $\sigma$).
}  
\label{Mvir_Mco} 
\end{figure*}

\begin{figure*}
\epsscale{1.0}
\plotone{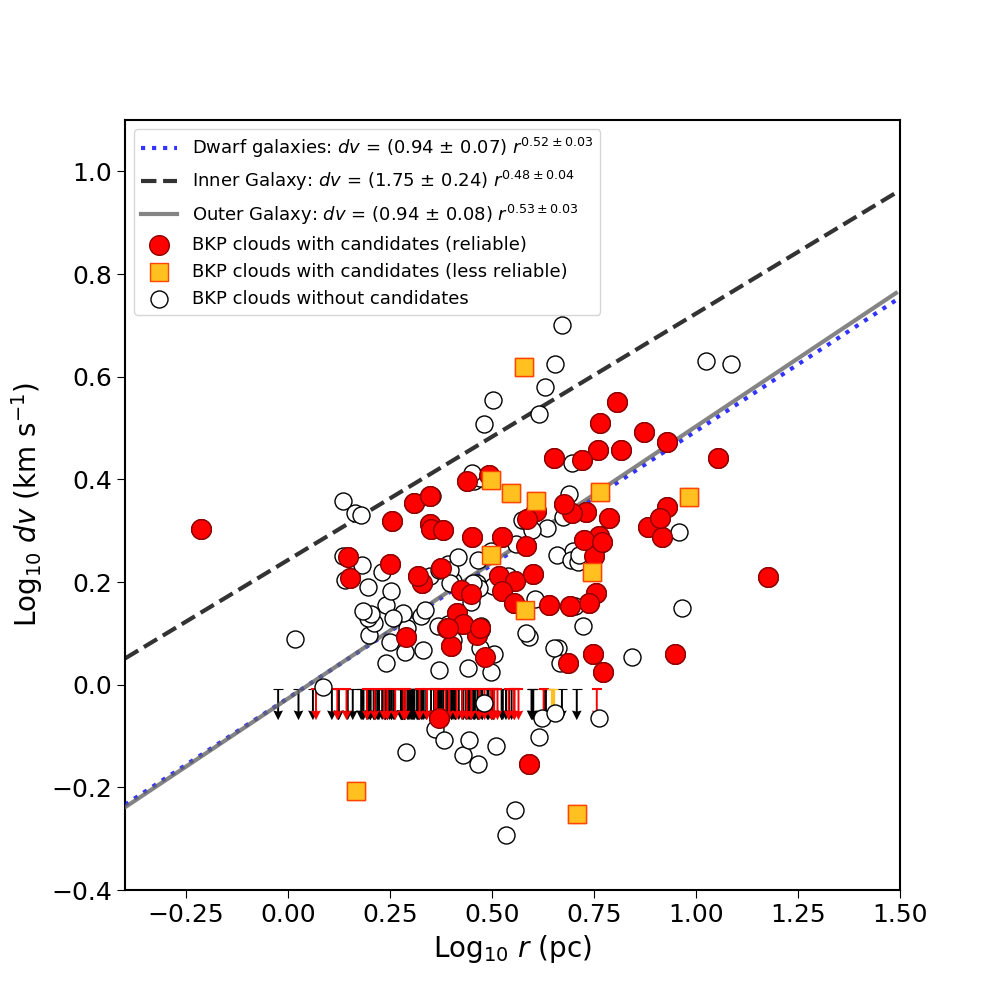}
\caption{
Size-linewidth relation of all BKP clouds in the outer Galaxy.
Black open circles show clouds without associated star-forming regions.
Red circles and yellow squares show clouds with associated candidate star-forming regions with a contamination rate of their parental clouds
of $<$ 30 \% (reliable) and $\ge$ 30 \% (less reliable), respectively.
The arrows indicate the upper limit ($dv$ = 0.98 km s$^{-1}$) for 277 clouds for which the linewidths are not derived in the BKP catalog.
The gray solid and black dashed lines show the result of the least-squares fitting
through the outer Galaxy data ($R_{\rm G}$ $\gtrsim$ 16 kpc):
$dv$ = (0.94 $\pm$ 0.08)$r^{0.53 \pm 0.03}$
and the inner Galaxy data: $dv$ = (1.75 $\pm$ 0.24)$r^{0.48 \pm 0.04}$ from \citet{Brand1995}, respectively.
The blue dotted line shows the result of the least-squares fitting through data of dwarf galaxies including SMC, LMC, and Wolf-Lundmark-Melotte (WLM):
$dv$ = (0.94 $\pm$ 0.07)$r^{0.52 \pm 0.03}$
from \citet{Rubio2015}.
} 
\label{Larson} 
\end{figure*}

\begin{figure*}
\epsscale{1.0}
\plotone{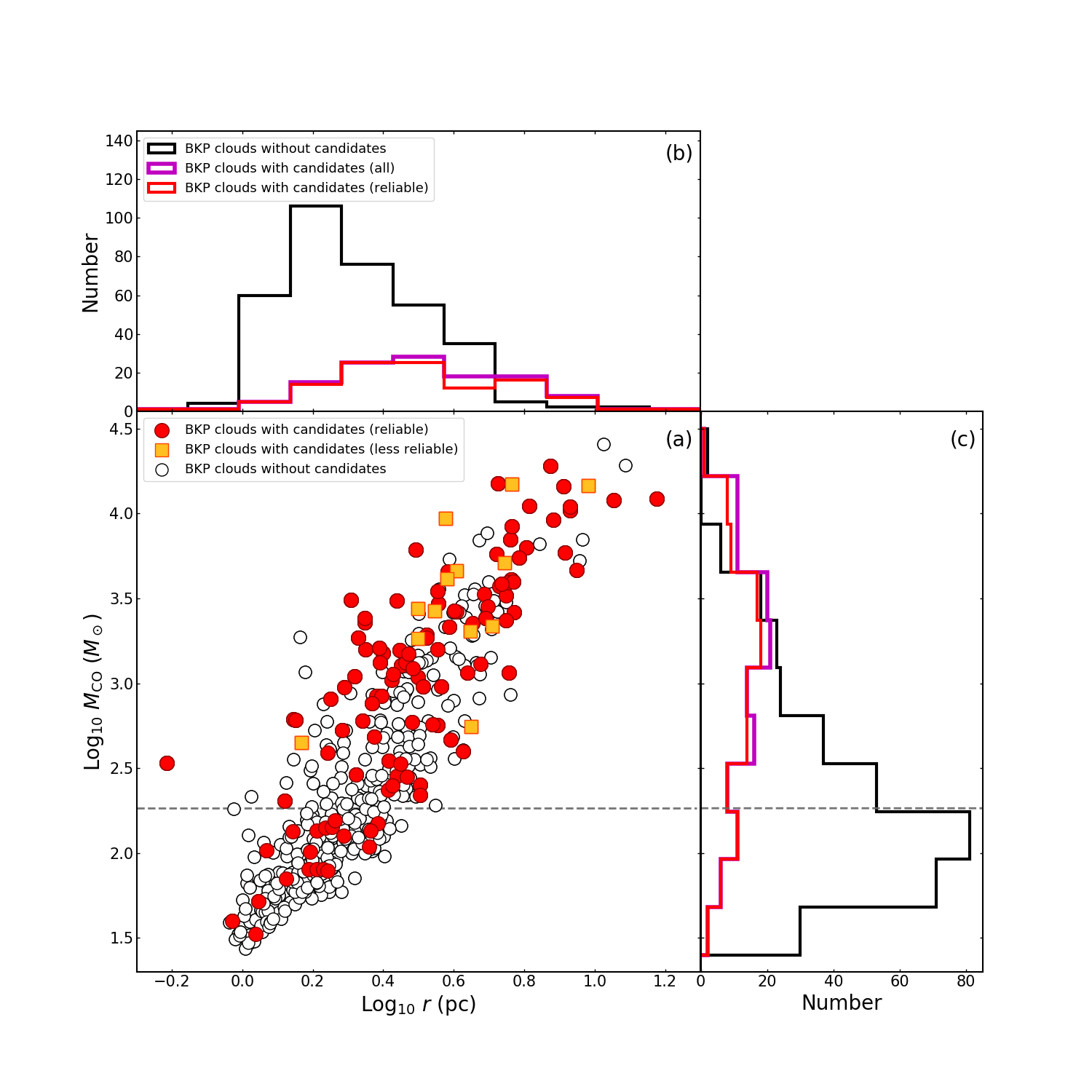}
\caption{
(a): Relation between cloud mass derived from CO intensity ($M_{\rm CO}$) and cloud size (radius; $r$) of BKP clouds in the outer Galaxy.
Red circles and yellow squares show clouds with associated candidate star-forming regions with a contamination rate of their parental clouds of
$<$ 30 \% (reliable) and $\ge$ 30 \% (less reliable), respectively.
Black open circles show clouds without associated candidates.
The gray-dashed line shows the mass threshold (183.6 $M_\odot$) for comparison of cloud properties at various distances up to $R_{\rm G}$ = 20.0 kpc.
(b): Number distribution of cloud size ($r$).
(c): Number distribution of cloud mass ($M_{\rm CO}$).
The black histograms show the number distribution of clouds without candidates.
The magenta and red histograms are the number distribution of clouds with all associated candidates and with only reliable associated candidates, respectively.
}  
\label{Mco_r} 
\end{figure*}

\begin{figure*}
\epsscale{1.0}
\plotone{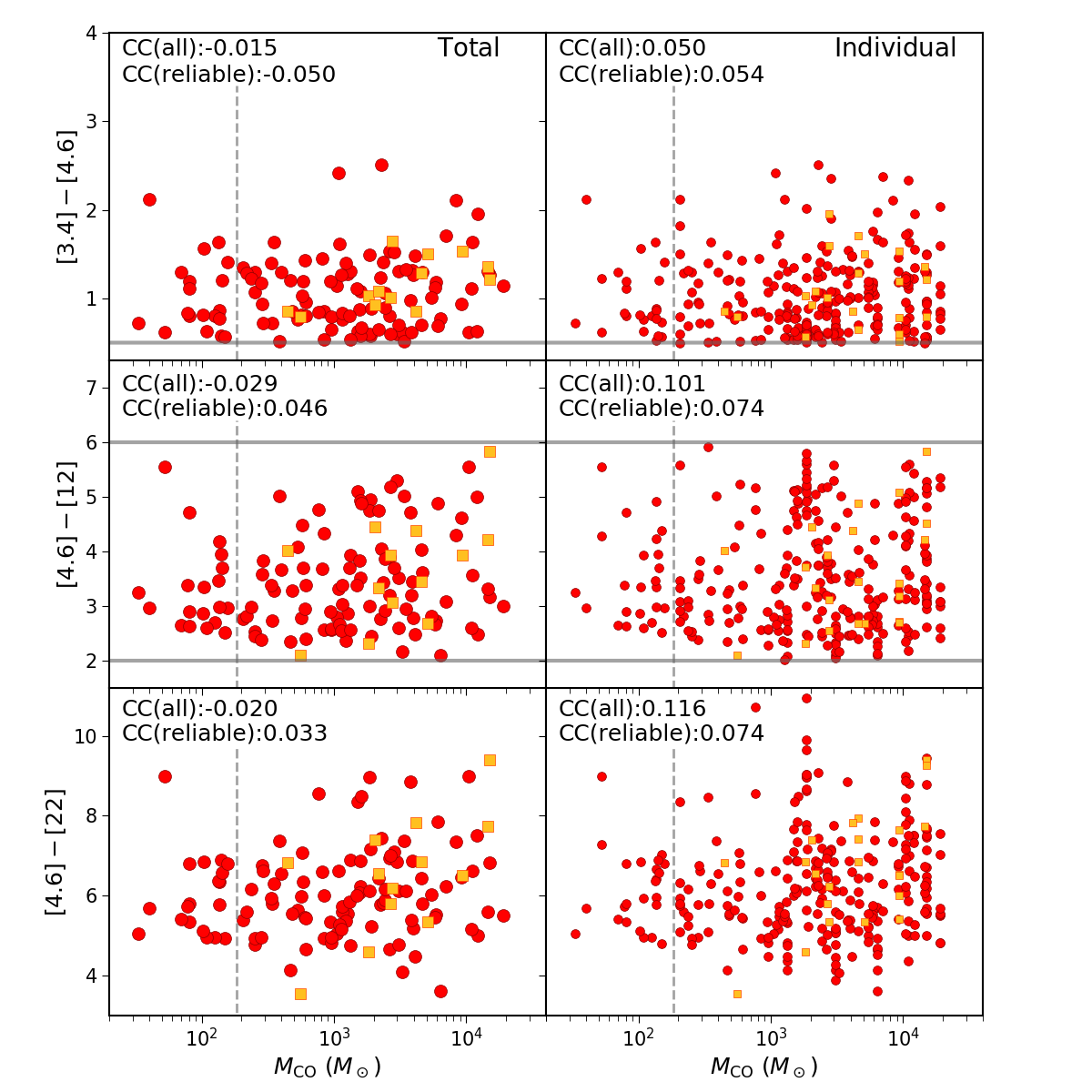}
\caption{
Left:
\textit{WISE} colors of 121 integrated (total) candidate star-forming regions in each parental cloud
(top: [3.4] - [4.6], middle: [4.6] - [12], bottom: [4.6] - [22]) plotted against cloud mass.
Red circles and yellow squares show clouds with associated star-forming regions with a contamination rate of their parental clouds
of $<$ 30 \% (reliable) and $\ge$ 30 \% (less reliable), respectively.
The gray lines show our identification criteria of star-forming region with \textit{WISE} data:
[3.4] $-$ [4.6] $\ge$ 0.5 and [4.6] $-$ [12] = 0.5 -- 2.0 (Paper I).
The gray-dashed lines show the completeness limit of cloud detection (183.6 M$_\odot$).
Correlation coefficients (CC) are noted in the top left corners.
Right: 
\textit{WISE} colors of 288 individual candidate star-forming regions
plotted against cloud mass.
Notations are the same as in the left panels.
} 
\label{Color-mass} 
\end{figure*}

\begin{figure*}
\epsscale{1.0}
\plotone{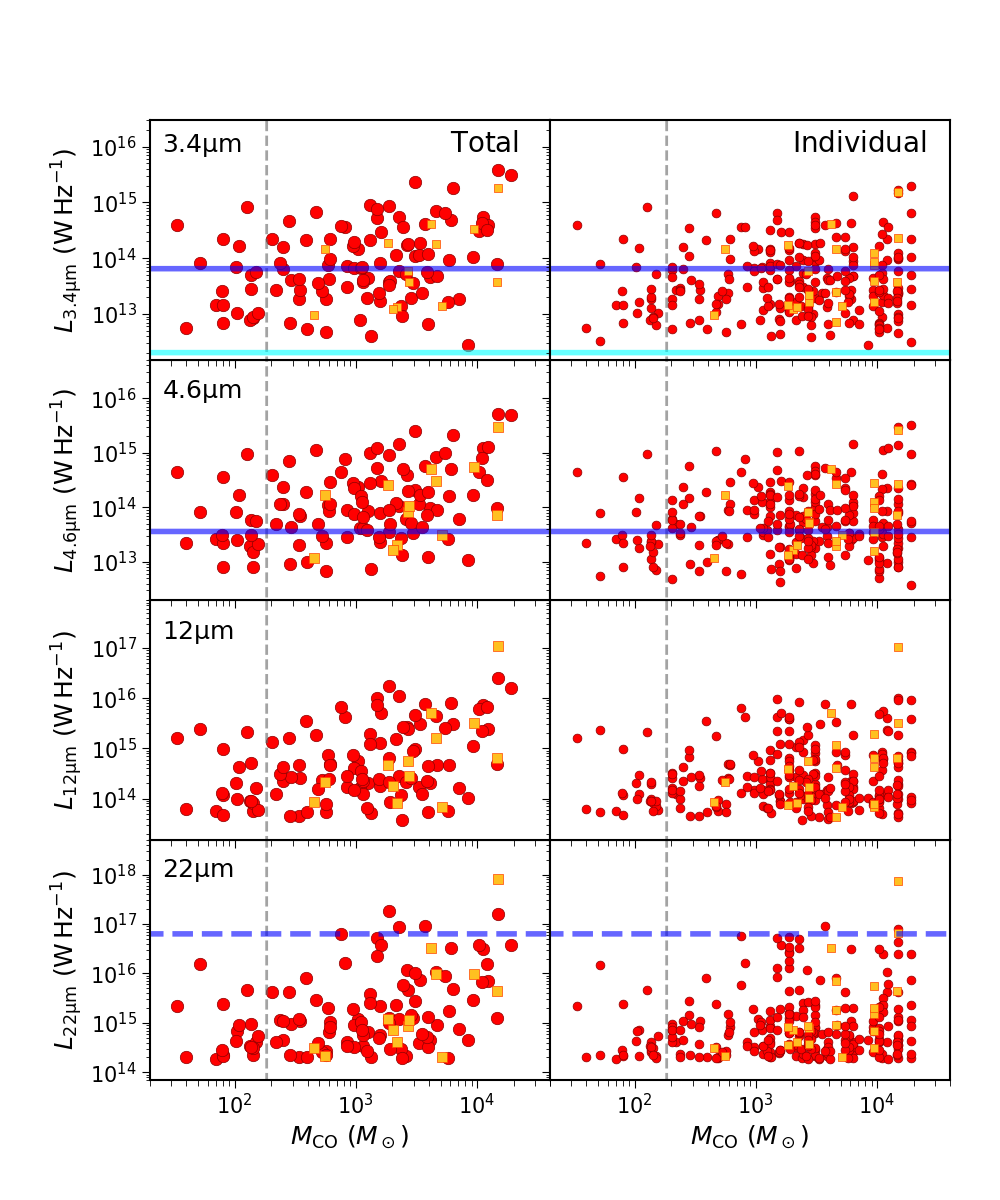}
\caption{
Left: Integrated (total) monochromatic luminosity of candidate star-forming regions in each cloud plotted against cloud mass.
Red circles and yellow squares show clouds with candidates with a contamination rate of their parental clouds
of $<$ 30 \% (reliable) and $\ge$ 30 \% (less reliable), respectively.
The cyan and blue lines show the flux densities for A0 and B0 stars in the main sequence \citep{Cox2000}.
Note that we used intrinsic $V$ $-$ $L$ colors (V band: 0.5555 $\micron$; L band: 3.547 $\micron$) and the absolute $V$-band magnitudes of A0 and B0 stars for calculating the apparent magnitudes of those stars for any of the four bands, since infrared colors, such as [3.4]$-$[22], are negligible for those early-type stars.
The blue-dashed line shows the 22 $\mu$m flux density for an H{\footnotesize II} regions ionized by B0 star \citep{Anderson2014}.
The gray-dashed lines show the completeness limit of cloud detection (183.6 M$_\odot$).
Right: Individual monochromatic luminosity of candidate star-forming regions plotted against cloud mass.
Notations are the same as in the left panels.
} 
\label{Mag-mass} 
\end{figure*}

\begin{figure*}
\epsscale{1.0}
\plotone{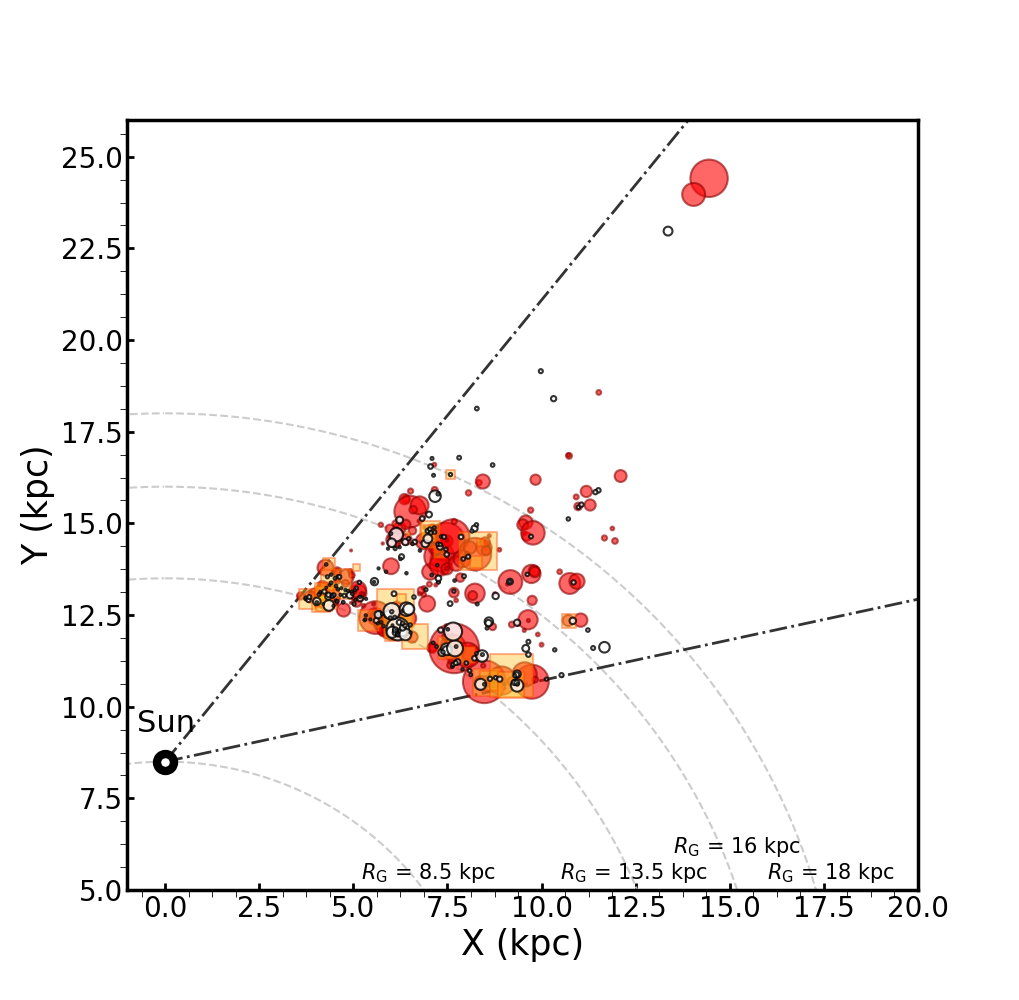}
\caption{
Distribution of BKP clouds beyond $R_{\rm G}$ of 13.5 kpc.
Red circles and yellow squares show clouds with associated candidate star-forming regions with a contamination rate of their parental clouds of $<$ 30 \% (reliable) and $\ge$ 30 \% (less reliable), respectively. Black open circles show clouds without associated candidates.
The size of these markers indicates the cloud mass (larger size corresponds to the larger mass).
The black-dotdashed lines indicate the area of the FCRAO survey (102$^\circ$.49 $\le$ $l$ $\le$ 141$^\circ$.54).
The gray-dotted curves indicate the $R_{\rm G}$ of 8.5, 13.5, 16.0, and 18.0 kpc.
Note that the two molecular clouds with kinematic distances of more than 20 kpc ($R_{\rm G}$ $>$ 27 kpc) are known to be actually located at a distance of 12 kpc ($R_{\rm G}$ = 19 kpc) from high-resolution optical spectra \citep[e.g.,][]{Smartt1996,Kobayashi2008}.
} 
\label{XY} 
\end{figure*}

\begin{figure*}
\epsscale{1.0}
\plotone{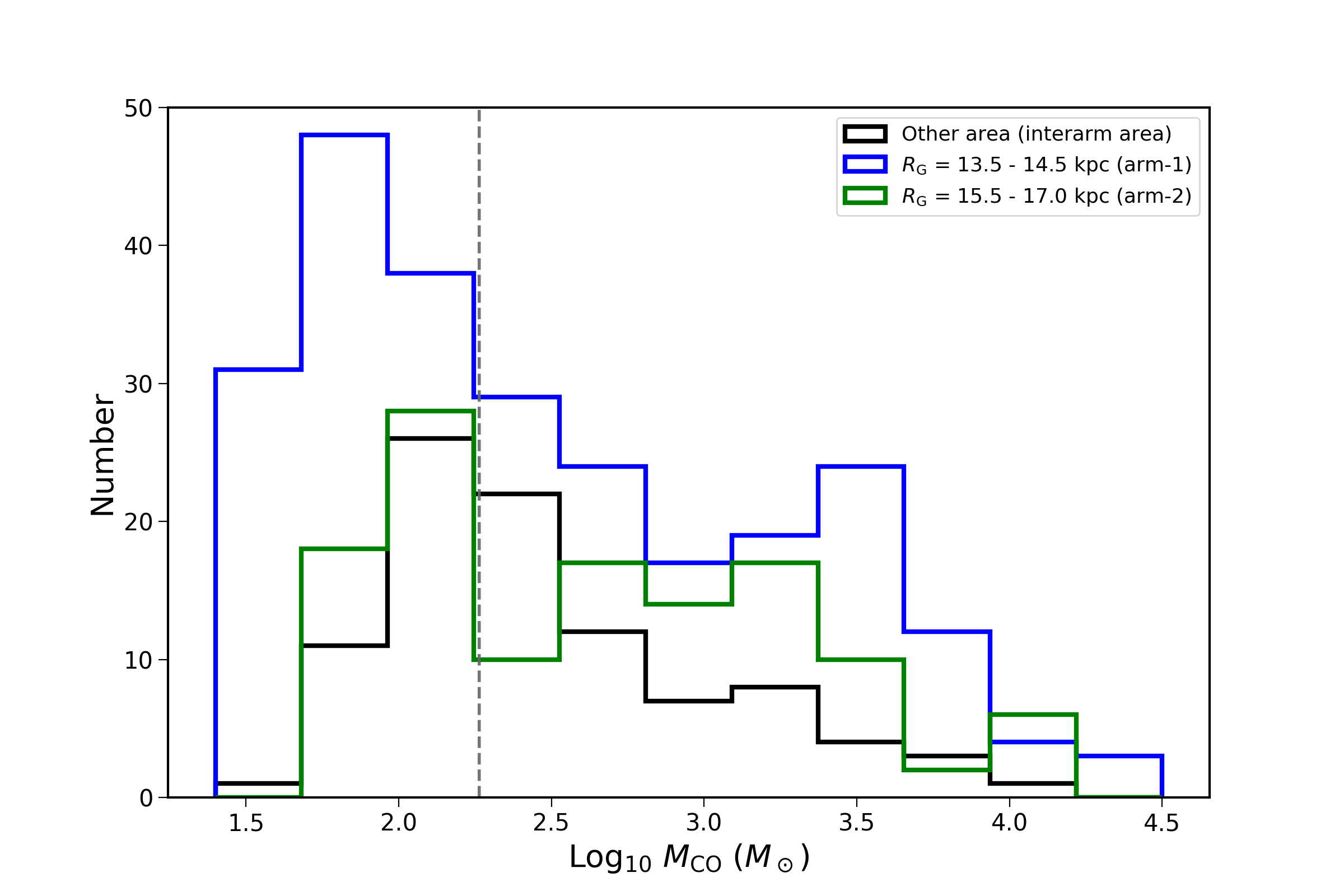}
\caption{
Number distributions of cloud mass ($M_{\rm CO}$).
The blue and green histograms show the number distribution of clouds 
in the range
$R_{\rm G}$ = 13.5--14.5 kpc (arm-1) and 15.5--17.0 kpc (arm-2), respectively.
The black histogram shows the number distribution of clouds in the other area
(interarm area).
The gray-dashed line shows the mass threshold (183.6 $M_\odot$) for comparison of cloud properties at various distances up to $R_{\rm G}$ = 20.0 kpc.
}
\label{hist-arm} 
\end{figure*}

\begin{figure*}
\epsscale{1.0}
\plotone{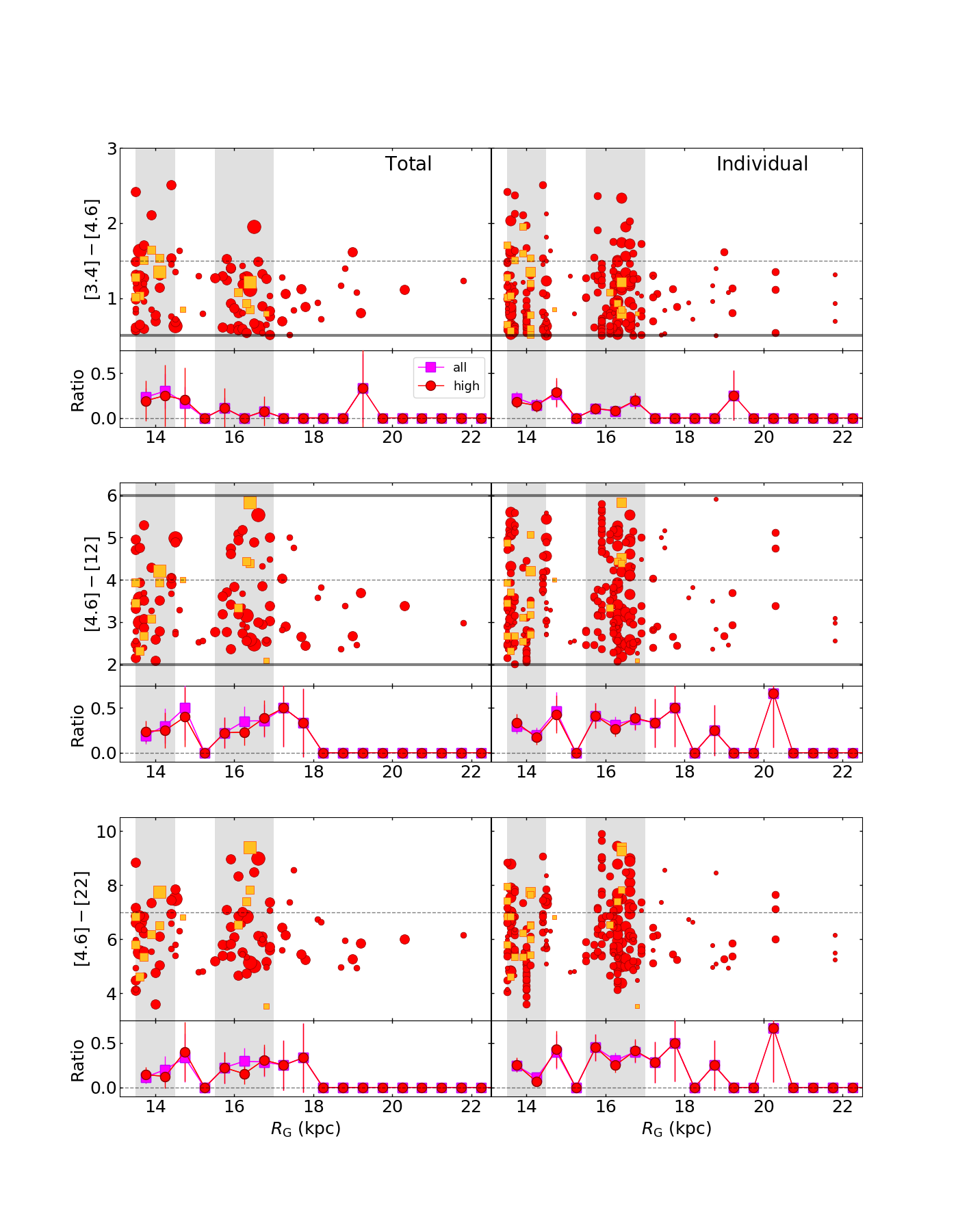}
\caption{
Left: Integrated (total) colors of candidate star-forming regions in each cloud (top: [3.4] $-$ [4.6], middle: [4.6] $-$ [12], bottom: [4.6] $-$ [22]) plotted against Galactocentric radius ($R_{\rm G}$).
Red circles and yellow squares show clouds with candidates with a contamination rate of their parental clouds
of $<$ 30 \% (reliable) and $\ge$ 30 \% (less reliable), respectively.
The size of these markers indicates the mass of their parental molecular clouds (small: $M_\odot$ $\le$ $M_{\rm CO}$ $<$ 10$^3$ $M_\odot$,
middle: 10$^3$ $M_\odot$ $\le$ $M_{\rm CO}$ $<$ 10$^4$ $M_\odot$,
large: 10$^4$ $M_\odot$ $\le$ $M_{\rm CO}$).
The gray lines show our identification criteria of candidates with \textit{WISE} data:
[3.4] $-$ [4.6] $\ge$ 0.5 and [4.6] $-$ [12] = 0.5 -- 2.0 (Paper I).
The number ratio of the reddened candidates (top: [3.4] $-$ [4.6] $>$ 1.5, middle: [4.6] $-$ [12] $>$ 4.0, bottom: [4.6] $-$ [22] $>$ 7.0) per total star-forming regions
is presented at the bottom of each color distributions
(magenta: all candidates, red: only candidates with contamination rate of $<$ 30 \%).
The error bars represent Poisson errors (1 $\sigma$).
The gray-dotted lines show the threshold for the reddened candidates.
The gray areas indicate the concentrated areas considered to be spiral arms.
Right: Individual colors of candidate star-forming regions plotted against cloud mass.
Notations are the same as in the left panels.
} 
\label{Color-rg}
\end{figure*}

\begin{figure*}
\epsscale{1.0}
\plotone{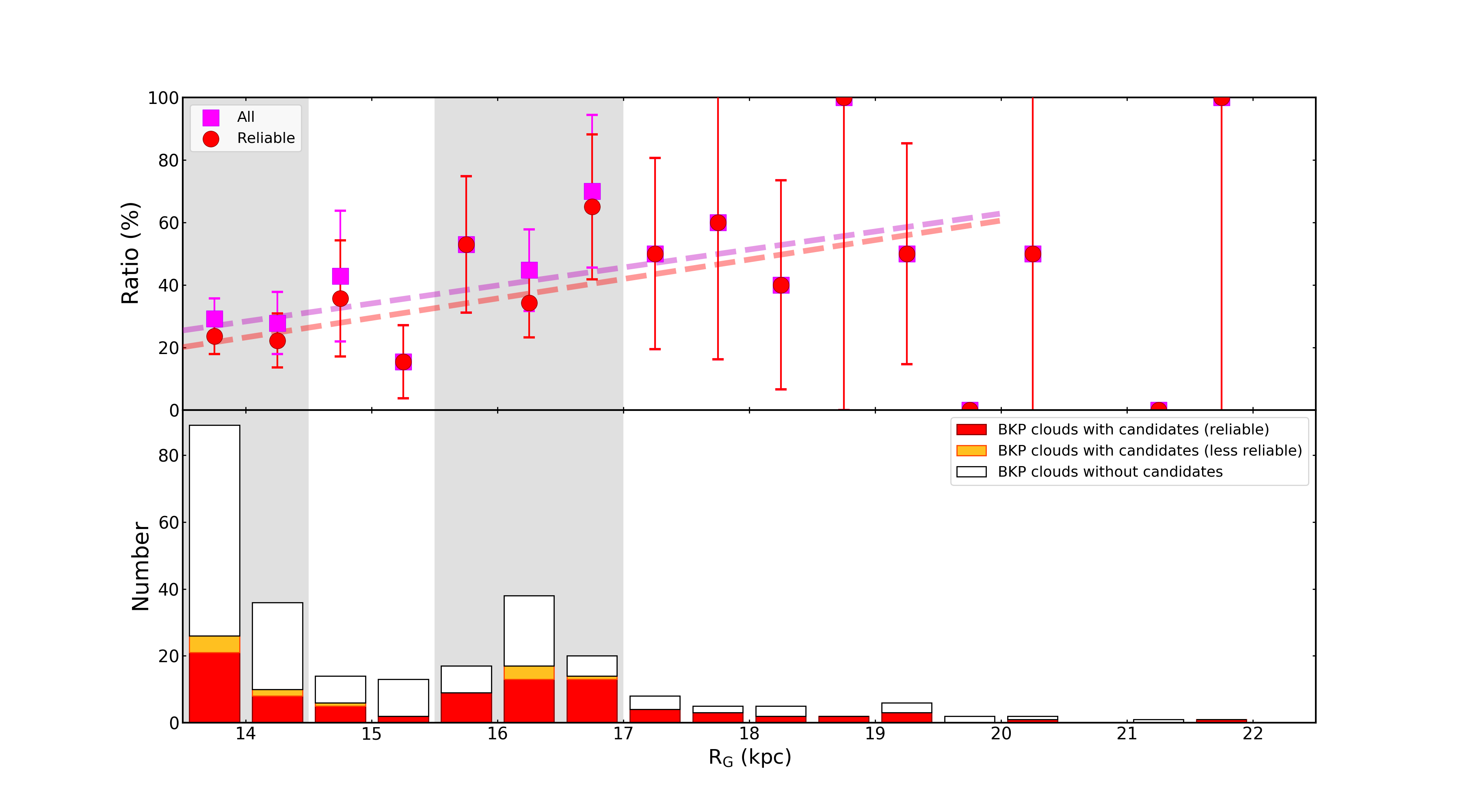}
\caption{
Top: Galactocentric radius variation of fraction of molecular clouds with candidate star-forming regions ($N_{\rm SF}$/$N_{\rm all}$) 
for clouds with $M_{\rm CO}$ $\ge$ 183.6 $M_\odot$.
The magenta squares and red circles show all the clouds with candidates and only clouds with reliable  candidates, respectively.
The error bars represent Poisson errors (1 $\sigma$).
The magenta and red dotted lines show the result of least-squares fittings:
$N_{\rm SF}$/$N_{\rm all}$ = 5.8($\pm$2.5)$R_{\rm G}$ - 52.3($\pm$36.5) (magenta: all clouds with candidates)
and $N_{\rm SF}$/$N_{\rm all}$ = 6.2($\pm$2.1)$R_{\rm G}$ - 64.0($\pm$31.0) (red: only clouds with reliable candidates), respectively.
Bottom: Galactocentric radius variation of the number clouds with and without candidates.
The red and orange bars show the number of clouds with  candidates with contamination rate of $<$ 30 \% (reliable) and $\ge$ 30 \% (less reliable), respectively.
The white bars show the number of clouds without  candidates.
Only clouds with $M_{\rm CO}$ $\ge$ 183.6 $M_\odot$ are plotted.
The gray areas indicate the concentrated areas considered to be spiral arms.
} 
\label{Total_rate_rg}
\end{figure*}

\begin{figure*}
\epsscale{1.0}
\plotone{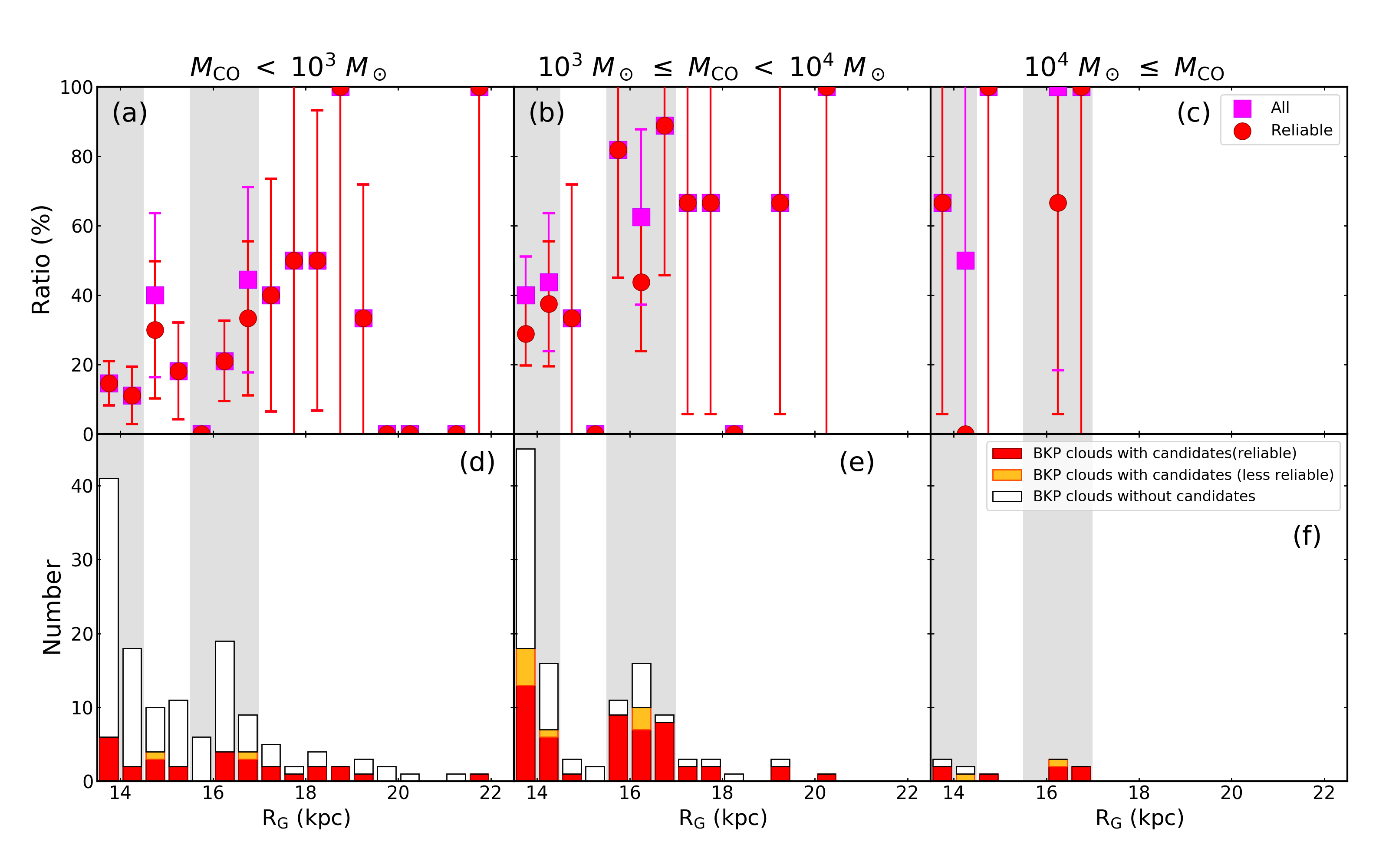}
\caption{
Same plot as Figure \ref{Total_rate_rg} but in three cloud mass ranges:
$M_{\rm CO}$ $<$ 10$^3$ $M_\odot$ (a,d), 
10$^3$ $M_\odot$ $\le$ $M_{\rm CO}$ $<$ 10$^4$ $M_\odot$ (b,e),
and 10$^4$ $M_\odot$ $\le$ $M_{\rm CO}$ (c,f).
Only clouds with $M_{\rm CO}$ $\ge$ 183.6 $M_\odot$ are plotted in (a,d).
The gray areas indicate the concentrated areas considered to be spiral arms.
} 
\label{Mass_rate_rg}
\end{figure*}

\begin{figure*}
\epsscale{1.0}
\plotone{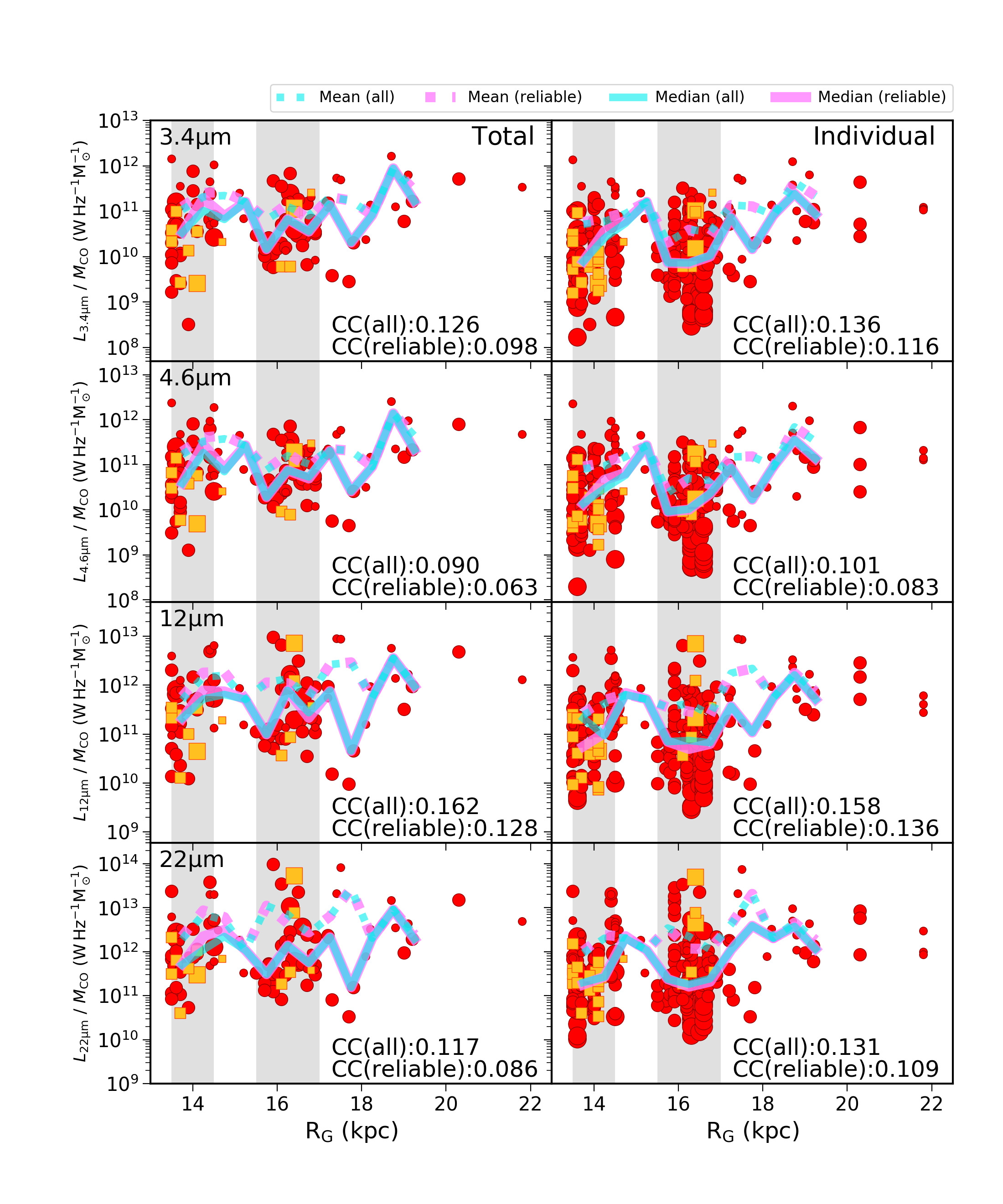}
\caption{
Left: Galactocentric radius variation of the integrated (total) monochromatic MIR luminosities of candidate star-forming regions per parental cloud mass ($L_{\rm MIR}$/$M_{\rm CO}$).
Red circles and yellow squares show clouds with candidates with a contamination rate of their parental clouds
of $<$ 30 \% (reliable) and $\ge$ 30 \% (less reliable), respectively.
The size of these markers indicates the mass of their parental molecular clouds (small: $M_{\rm CO}$ $<$ 10$^3$ $M_\odot$,
middle: 10$^3$ $M_\odot$ $\le$ $M_{\rm CO}$ $<$ 10$^4$ $M_\odot$,
large: 10$^4$ $M_\odot$ $\le$ $M_{\rm CO}$).
Magenta solid and dotted lines indicate median and mean values of only candidates with a contamination rate of less than 30 \% (reliable), respectively.
Cyan solid and dotted lines indicate median and mean values of all candidates.
These values are calculated in a 0.5 kpc binning from $R_{\rm G}$ of 13.5 to 20.0 kpc.
The gray areas indicate the concentrated areas considered to be spiral arms.
Correlation coefficients (CC) are noted in the bottom right corners.
Right: Galactocentric radius variation of the individual monochromatic MIR luminosities of candidate star-forming regions per parental cloud mass ($L_{\rm MIR}$/$M_{\rm CO}$).
Notations are the same as in the left panels.
} 
\label{Lumimass_rg}
\end{figure*}

We report the properties of newly identified candidate star-forming regions in the outer Galaxy with \textit{WISE} MIR and FCRAO CO survey data.
The main results are as follows, and trends with $R_{\rm G}$ are summarized in Table 1:

\begin{enumerate}
    \item There are differences between the properties of molecular clouds with and without candidate star-forming regions:
    1) The slope of the mass spectrum of molecular clouds without candidates is steeper ($\alpha$ = -1.57 $\pm$ 0.08) than that with candidates ($\alpha$ = -1.04 $\pm$ 0.03). 
    2) Almost all clouds with candidates are bound by self-gravity, while those without candidates are not bound.
    3) The column density of molecular clouds with candidates is larger than twice
    without candidates
    (Section \ref{sec:3}; Figures \ref{Mass_spec}--\ref{Mvir_Mco} and \ref{Mco_r}). 
    
    \item There is no correlation between the MIR color ([3.4] - [4.6], [4.6] - [12], and [4.6] - [22]) of the candidates in low mass ($\le$ 10$^3$ $M_\odot$) and high-mass ($>$ 10$^3$ $M_\odot$) molecular clouds.
    Candidates with brighter luminosity ($L_{\rm 3.4 \micron}$ $\ge$ 10$^{15}$ W Hz$^{-1}$,
    $L_{\rm 4.6 \micron}$ $\ge$ 10$^{15}$ W Hz$^{-1}$,
    $L_{\rm 12 \micron}$ $\ge$ 10$^{16}$ W Hz$^{-1}$,
    and $L_{\rm 22 \micron}$ $\ge$ 10$^{17}$ W Hz$^{-1}$) are only associated with the high-mass molecular clouds ($>$ 10$^3$ $M_\odot$).
    The threshold of $L_{\rm 22 \micron}$ roughly corresponds to the luminosity of
    of an \ion{H}{2} region ionized by a B0 star \citep[blue dotted lines in Figure \ref{Mag-mass};][]{Anderson2014}
    (Section \ref{sec:4}; Figures \ref{Color-mass} and \ref{Mag-mass}).
    
    \item Candidates with redder color ([3.4] - [4.6] $>$ 1.5, [4.6] - [12] $>$ 4.0, and [4.6] - [22] $>$ 7.0) are mostly located at
    $R_{\rm G}$ $<$ 18 kpc. This blueing toward larger $R_{\rm G}$ could be due to the stochastic effect with the small number of sources.
    However, if the blueing trend is a real feature, it can be interpreted as the result of the absence of massive star-forming regions at $R_{\rm G}$ $>$ 18 kpc
    (Section \ref{sec:5_1}; Figure \ref{Color-rg}).
    
    \item The two SFE parameters converting from H$_2$ gas mass to stellar mass:
    1) the fraction of molecular clouds with candidates,
    and 2) the monochromatic MIR luminosities of the candidates per parental cloud mass,
    do not show a clear evidence of change from $R_{\rm G}$ = 13.5 to 20.0 kpc where a largely varying environment is expected.
    This suggests that the SFE per molecular cloud does not depend on the environmental parameters, such as metallicity and gas surface density
    (Sections \ref{sec:5_21}--\ref{sec:5_23}; Figures \ref{Total_rate_rg}--\ref{Lumimass_rg}).
    
    Previous studies reported that the SFE/yr converting from \ion{H}{1} gas mass to stellar mass decreases with increasing $R_{\rm G}$ in the outer Galaxy.
    Based on the assumption that SFE is proportional to the SFE/yr, our result suggests that the decrease of the SFE/yr is due to a decrease in conversion efficiency of \ion{H}{1} gas mass to H$_2$ gas mass. Previous studies also reported that SFE/yr converting from H$_2$ gas mass to stellar mass  is constant in the inner Galaxy, although $\Sigma _{\rm H_2}$ decreases with increasing $R_{\rm G}$. Therefore, our result also suggests that this trend also holds for the outer Galaxy
    (Section \ref{sec:5_23}).
\end{enumerate}

\begin{acknowledgments}
We thank the anonymous reviewer for a careful reading and thoughtful suggestions that significantly improved this paper.
We also would like to thank Hauyu Baobab Liu and Michihiro Takami for helpful discussions.
P.M.K. is supported by the Ministry of Science and Technology (MoST) through grants MoST 109-2112-M-001-022, MoST 110-2112-M-001-057, and MoST 111-2112-M-001-003.
This work is supported by JSPS KAKENHI Grant No. 18H05441.
This publication makes use of data products from the Wide-field Infrared Survey Explorer, which is a joint project of the University of California, Los Angeles, and the Jet Propulsion Laboratory/California Institute of Technology, funded by the National Aeronautics and Space Administration.
This research has made use of the NASA/IPAC Infrared Science Archive, which is funded by the National Aeronautics and Space Administration and operated by the California Institute of Technology.
This research made use of Montage. It is funded by the National Science Foundation under Grant Number ACI-1440620, and was previously funded by the National Aeronautics and Space Administration's Earth Science Technology Office, Computation Technologies Project, under Cooperative Agreement Number NCC5-626 between NASA and the California Institute of Technology.
The research presented in this paper has used data from the Canadian Galactic Plane Survey, a Canadian project with international partners, supported by the Natural Sciences and Engineering Research Council. 
This research used the facilities of the Canadian Astronomy Data Centre operated by the National Research Council of Canada with the support of the Canadian Space Agency. 
\end{acknowledgments}

\begin{deluxetable*}{cccccc}
\tablecaption{Star-Formation Features versus $R_{\rm G}$}
\tablehead{
\colhead{Feature} &
\multicolumn{4}{c}{$R_{\rm G}$} &
\colhead{Section}
\\
\cline{2-5} \colhead{} & \colhead{$\le$14.5 kpc} & \colhead{14.5--15.5 kpc} & \colhead{15.5--17.0 kpc} & \colhead{$\ge$17.0 kpc} & \colhead{}\\
\colhead{} & \colhead{(spiral arm)} & \colhead{(interarm)} & \colhead{(spiral arm)} & \colhead{(outside, interarm)}
}
\startdata
Gas (\ion{H}{1}, H$_2$) surface density  & \multicolumn{4}{c}{Decreasing with $R_{\rm G}$} & \ref{sec:1}\tablenotemark{a} \\
Metallicity  & \multicolumn{4}{c}{Decreasing with $R_{\rm G}$} & \ref{sec:1}\tablenotemark{b} \\ 
Massive clouds  & present & absent & present & absent & \ref{sec:2_12}, \ref{sec:5_new} \\
($M_{\rm CO}$ $\ge$ 10$^4$ $M_\odot$) \\
Bright MIR candidates
& present & absent & present & absent & \ref{sec:2_21}\\
($L_{22 \micron}$ $\ge$ 10$^{17}$ W Hz${^{-1}}$)\\
Red candidates
& present & only very few & present  & only very few   & \ref{sec:5_1} \\
(e.g., [3.4]-[4.6] $>$ 1.5]) & & & & (absent beyond 18 kpc) &\\
SFE parameter 1            & \multicolumn{4}{c}{No change with $R_{\rm G}$ (possible slight increase)} & \ref{sec:5_21} \\ 
($N_{\rm SF}$/$N_{\rm all}$) \\
SFE parameter 2            & \multicolumn{4}{c}{No change with $R_{\rm G}$} & \ref{sec:5_22} \\ 
($L_{\rm MIR}$/$M_{\rm CO}$)
\enddata
\tablenotetext{a}{References are \citet{Wolfire2003}, \citet{Heyer2015}, and \citet{Nakanishi2016}}
\tablenotetext{b}{References are \citet{Smartt1997} and \citet{Fernandez2017}}
\label{tab:results}
\end{deluxetable*}
\clearpage
\appendix
\section{Threshold values}\label{sec:a_1}
\begin{figure*}
\epsscale{1.0}
\plotone{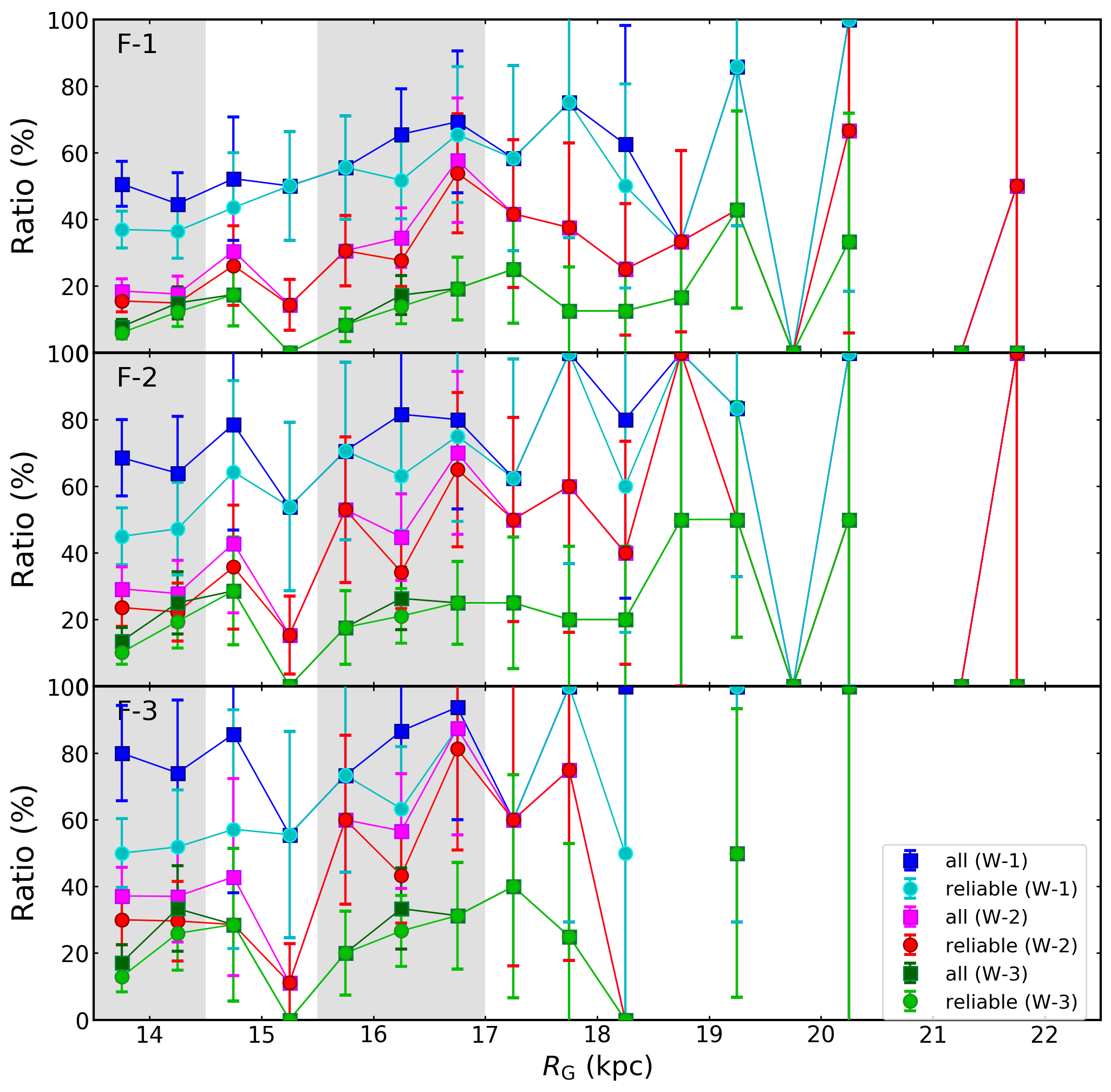}
\caption{
Galactocentric radius variation of $N_{\rm SF}$/$N_{\rm all}$
derived from three threshold values of each \textit{WISE} and FCRAO data.
The details of adopted threshold values (symbol name: W-1, W-2, W-3, F-1, F-2, and F-3) are described in Table 2.
The circles and squares show the ratio for clouds with associated candidates with a contamination rate less than 30 \% (reliable) and all clouds with associated candidates, respectively.
The error bars represent Poisson errors (1 $\sigma$).
The gray areas indicate the concentrated areas considered to be spiral arms.
} 
\label{Total_rate_rg_comp}
\end{figure*}

\begin{figure*}
\epsscale{1.0}
\plotone{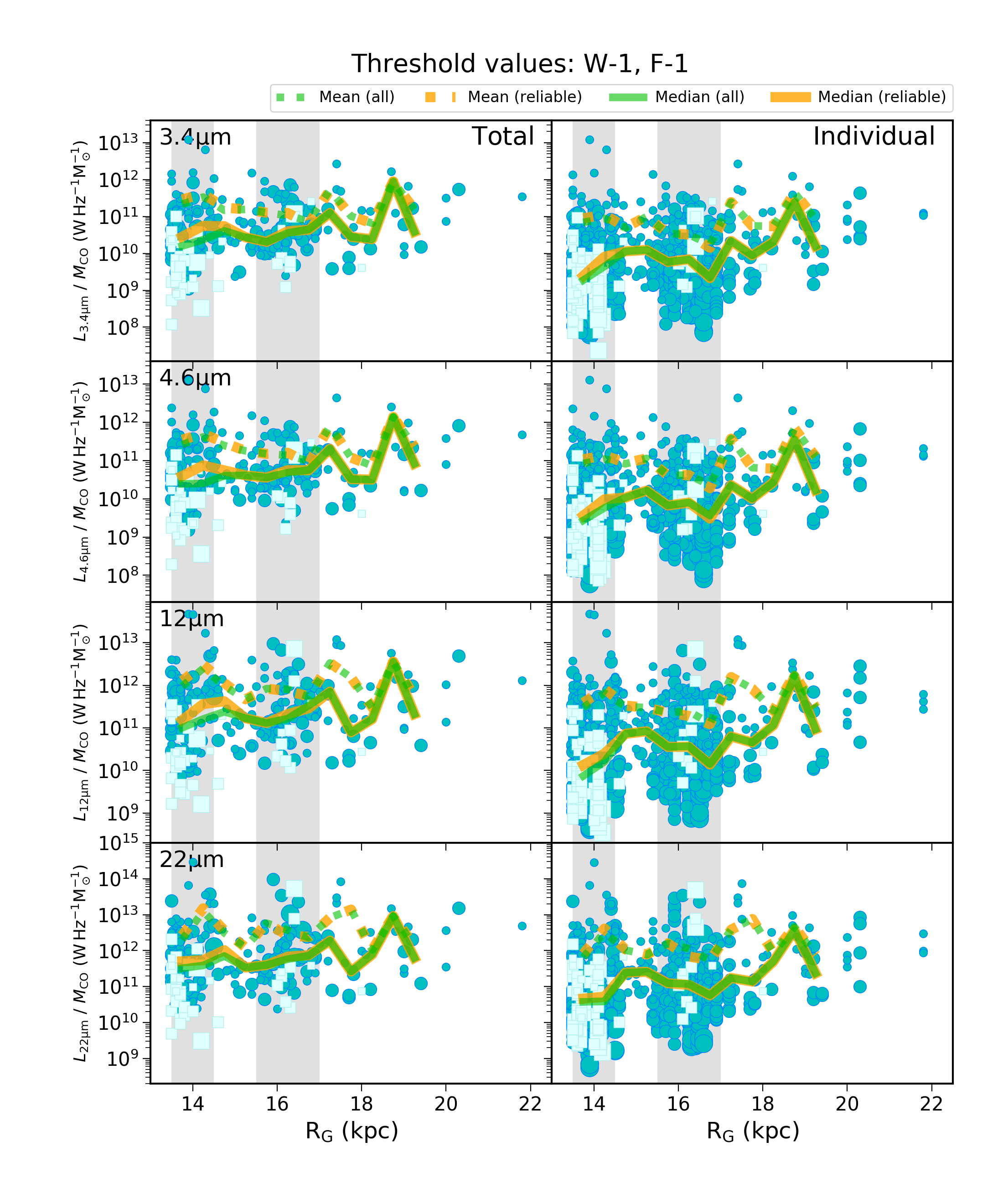}
\caption{
The total (left) and individual (right) $L_{\rm MIR}$/$M_{\rm CO}$
plotted against galactocentric radius.
The adopted completeness limit is W-1 and F-1 (see Table 2).
Cyan circles and light cyan squares show the star-forming regions with contamination rate $<$ and $\ge$ 30 \%, respectively.
The size of these markers indicates the mass of their parental molecular clouds
(small: $\le$ $M_{\rm CO}$ $<$ 10$^3$ $M_\odot$,
middle: 10$^3$ $M_\odot$ $\le$ $M_{\rm CO}$ $<$ 10$^4$ $M_\odot$,
large: 10$^4$ $M_\odot$ $\le$ $M_{\rm CO}$).
Orange solid and dotted lines indicate median and mean values of only candidates with a contamination rate of less than 30 \% (reliable), respectively.
Green solid and dotted lines indicate median and mean values of all candidates.
These values are calculated in a 0.5 kpc binning from $R_{\rm G}$ of 13.5 to 20.0 kpc.
The gray areas indicate the concentrated areas considered to be spiral arms.
} 
\label{Lumimass_CompMass-0_WISEComp-all_rg}
\end{figure*}

\begin{figure*}
\epsscale{1.0}
\plotone{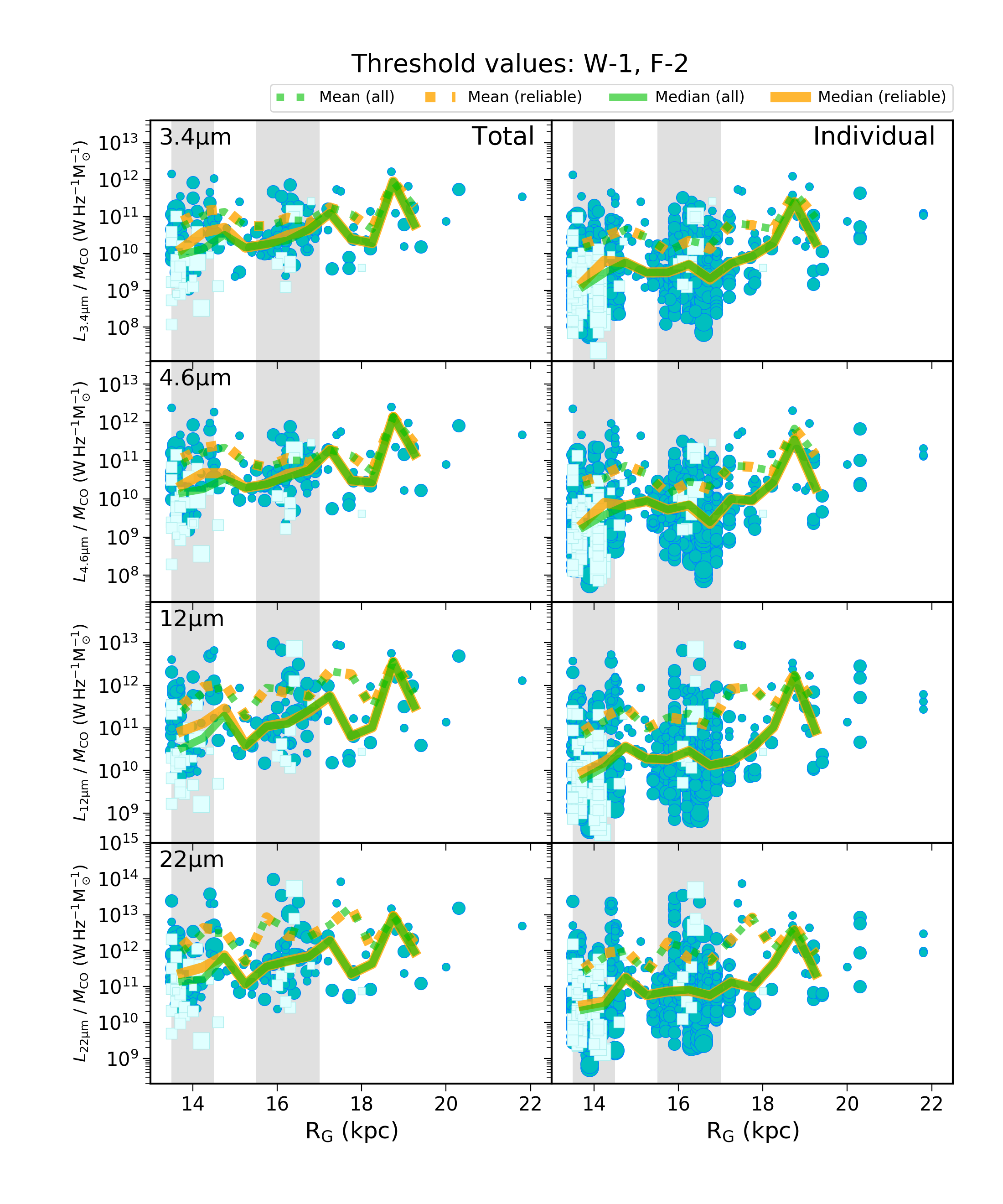}
\caption{
Identical to Figure \ref{Lumimass_CompMass-0_WISEComp-all_rg}, but for the threshold values W-1 and F-2.
} 
\label{Lumimass_CompMass-1_WISEComp-all_rg}
\end{figure*}

\begin{figure*}
\epsscale{1.0}
\plotone{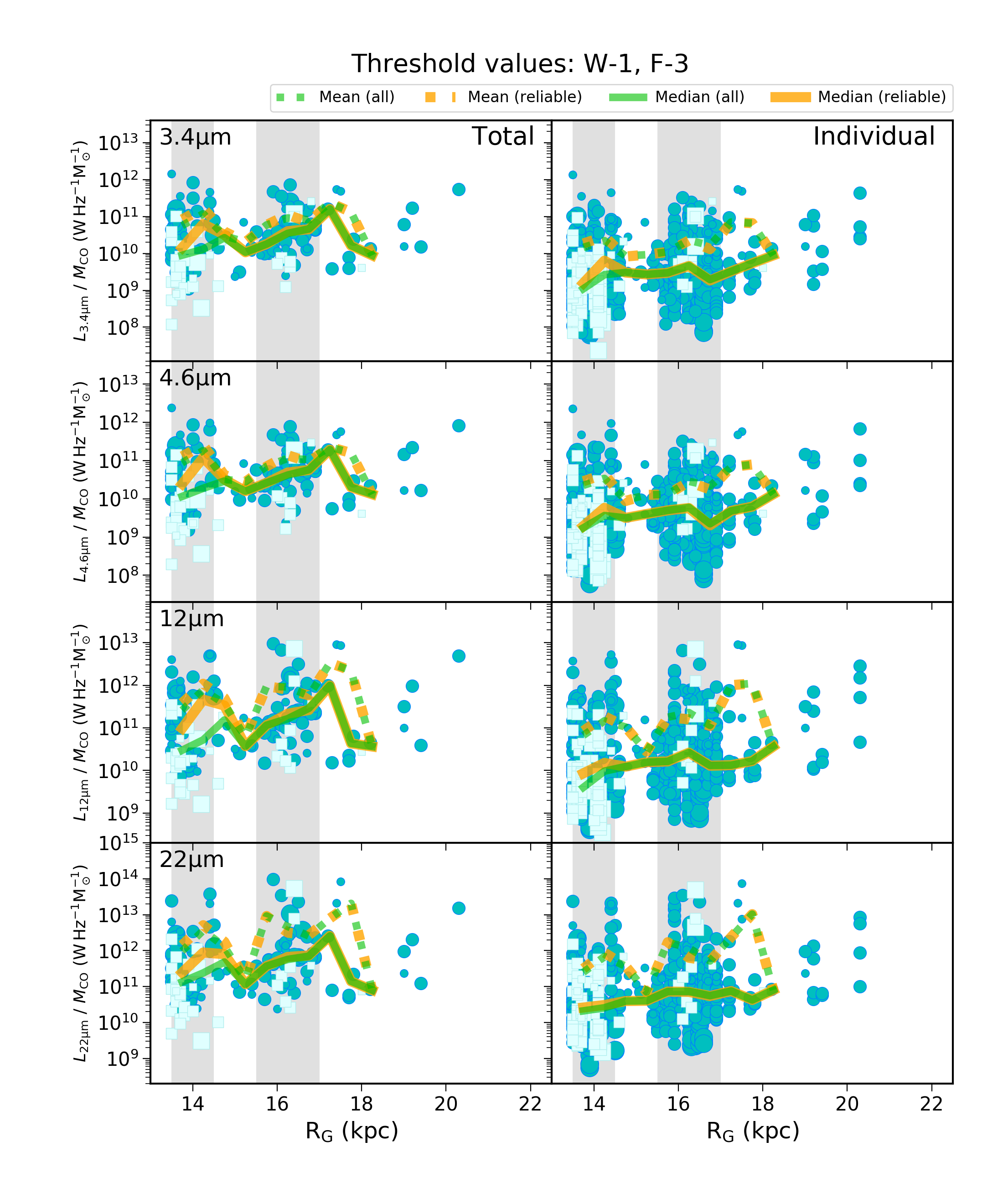}
\caption{
Identical to Figure \ref{Lumimass_CompMass-0_WISEComp-all_rg}, but for the threshold values W-1 and F-3.
} 
\label{Lumimass_CompMass-2_WISEComp-all_rg}
\end{figure*}

\begin{figure*}
\epsscale{1.0}
\plotone{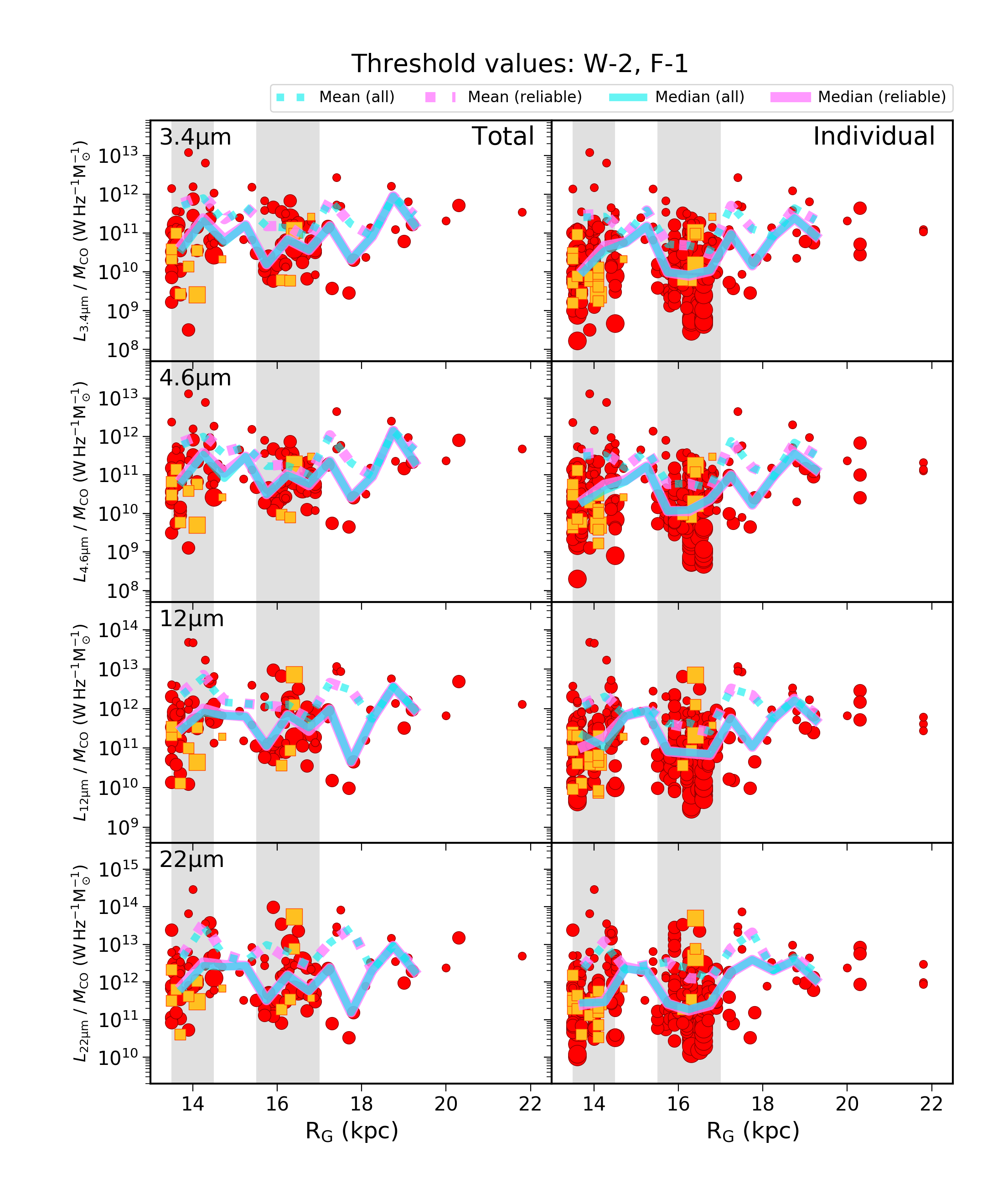}
\caption{
The total (left) and individual (right) $L_{\rm MIR}$/$M_{\rm CO}$
plotted against galactocentric radius.
The adopted completeness limit is W-2 and F-1 (see Table 2).
Red circles and yellow squares show the star-forming regions with contamination rate $<$ and $\ge$ 30 \%, respectively.
The size of these markers indicates the mass of their parental molecular clouds
(small: $\le$ $M_{\rm CO}$ $<$ 10$^3$ $M_\odot$,
middle: 10$^3$ $M_\odot$ $\le$ $M_{\rm CO}$ $<$ 10$^4$ $M_\odot$,
large: 10$^4$ $M_\odot$ $\le$ $M_{\rm CO}$).
Magenta solid and dotted lines indicate median and mean values of only candidates with a contamination rate of less than 30 \% (reliable), respectively.
Cyan solid and dotted lines indicate median and mean values of all candidates.
These values are calculated in a 0.5 kpc binning from $R_{\rm G}$ of 13.5 to 20.0 kpc.
The gray areas indicate the concentrated areas considered to be spiral arms.
} 
\label{Lumimass_CompMass-0_rg}
\end{figure*}

\begin{figure*}
\epsscale{1.0}
\plotone{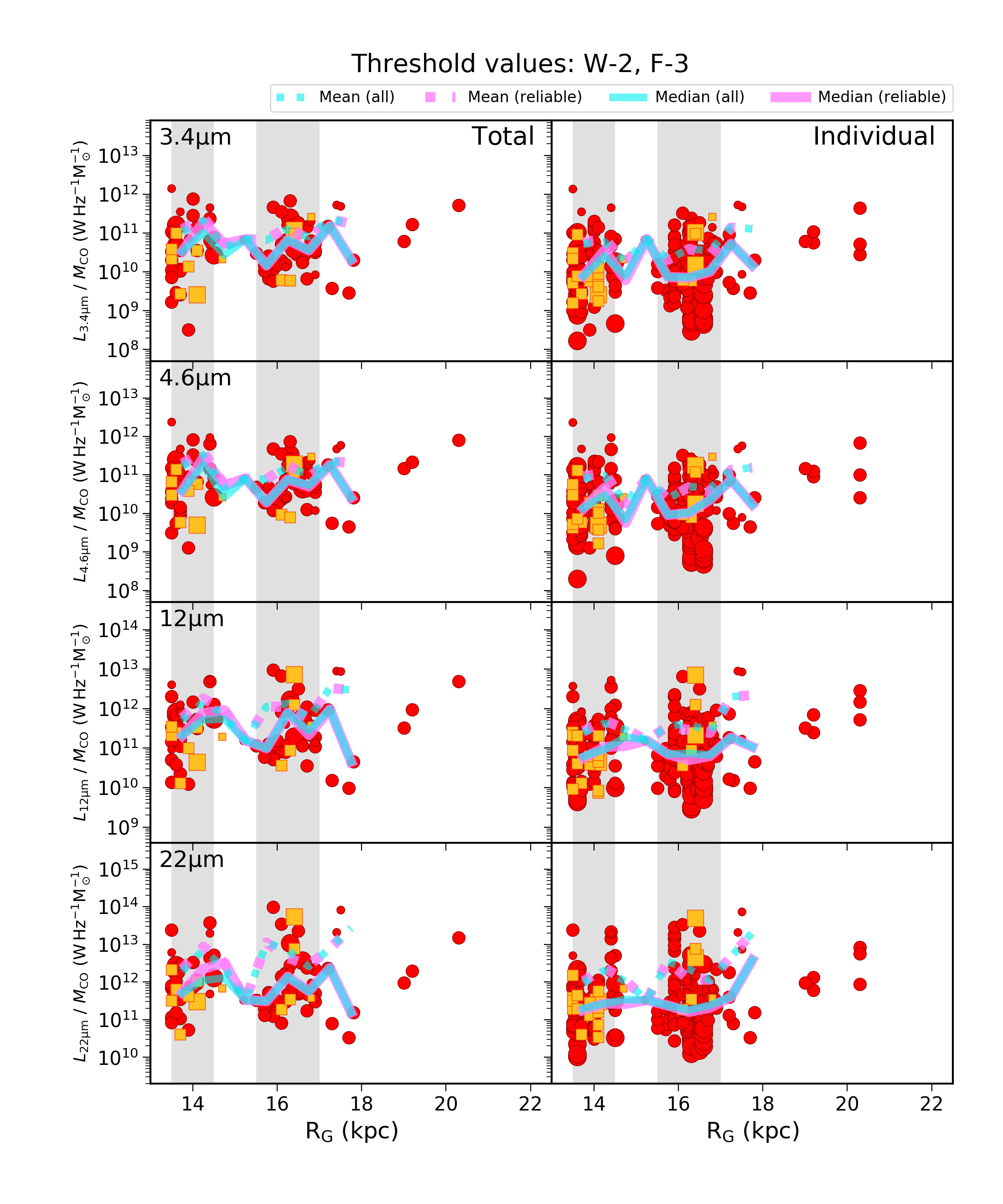}
\caption{
Identical to Figure \ref{Lumimass_CompMass-0_rg}, but for the threshold values W-2 and F-3.
} 
\label{Lumimass_CompMass-2_rg}
\end{figure*}

\begin{figure*}
\epsscale{1.0}
\plotone{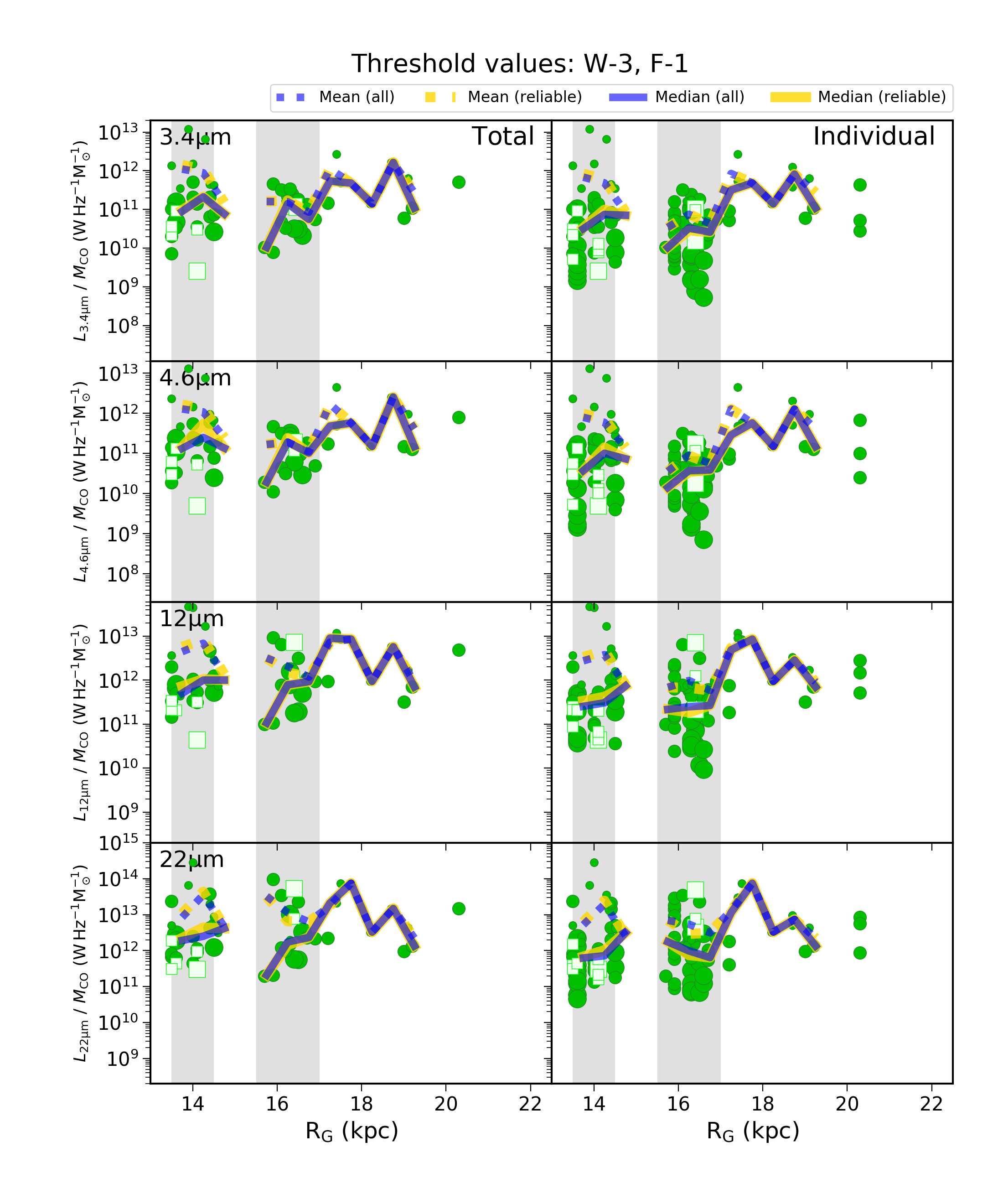}
\caption{
The total (left) and individual (right) $L_{\rm MIR}$/$M_{\rm CO}$
plotted against galactocentric radius.
The adopted completeness limit is W-3 and F-1 (see Table 2).
Green circles and right green squares show the star-forming regions with contamination rate $<$ and $\ge$ 30 \%, respectively.
The size of these markers indicates the mass of their parental molecular clouds
(small: $\le$ $M_{\rm CO}$ $<$ 10$^3$ $M_\odot$,
middle: 10$^3$ $M_\odot$ $\le$ $M_{\rm CO}$ $<$ 10$^4$ $M_\odot$,
large: 10$^4$ $M_\odot$ $\le$ $M_{\rm CO}$).
Yellow solid and dotted lines indicate median and mean values of only candidates with a contamination rate of less than 30 \% (reliable), respectively.
Blue solid and dotted lines indicate median and mean values of all candidates.
These values are calculated in a 0.5 kpc binning from $R_{\rm G}$ of 13.5 to 20.0 kpc.
The gray areas indicate the concentrated areas considered to be spiral arms.
} 
\label{Lumimass_CompMass-0_WISEComp-m2_rg}
\end{figure*}

\begin{figure*}
\epsscale{1.0}
\plotone{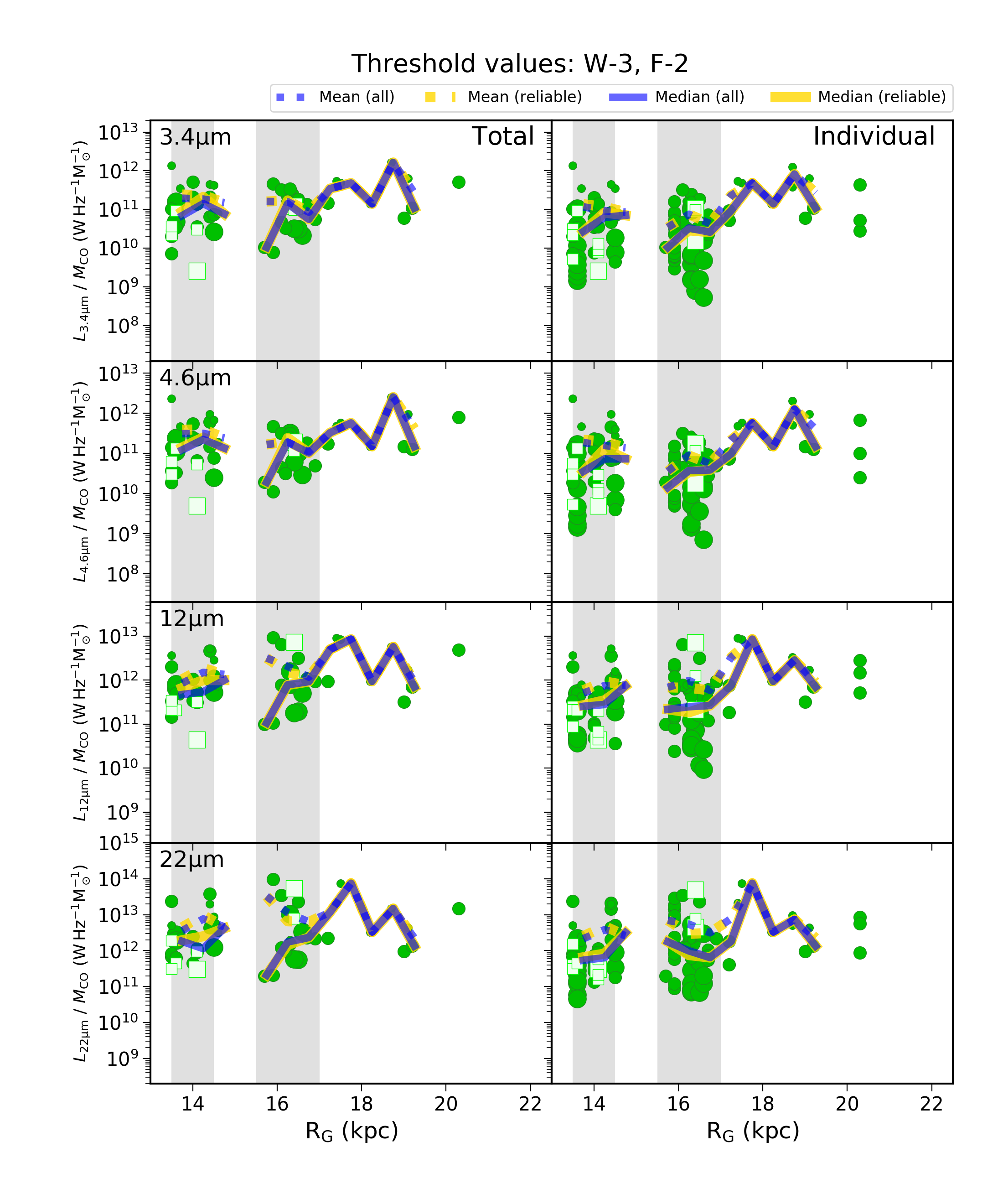}
\caption{
Identical to Figure \ref{Lumimass_CompMass-0_WISEComp-m2_rg}, but for the threshold values W-3 and F-2.}
\label{Lumimass_CompMass-1_WISEComp-m2_rg}
\end{figure*}

\begin{figure*}
\epsscale{1.0}
\plotone{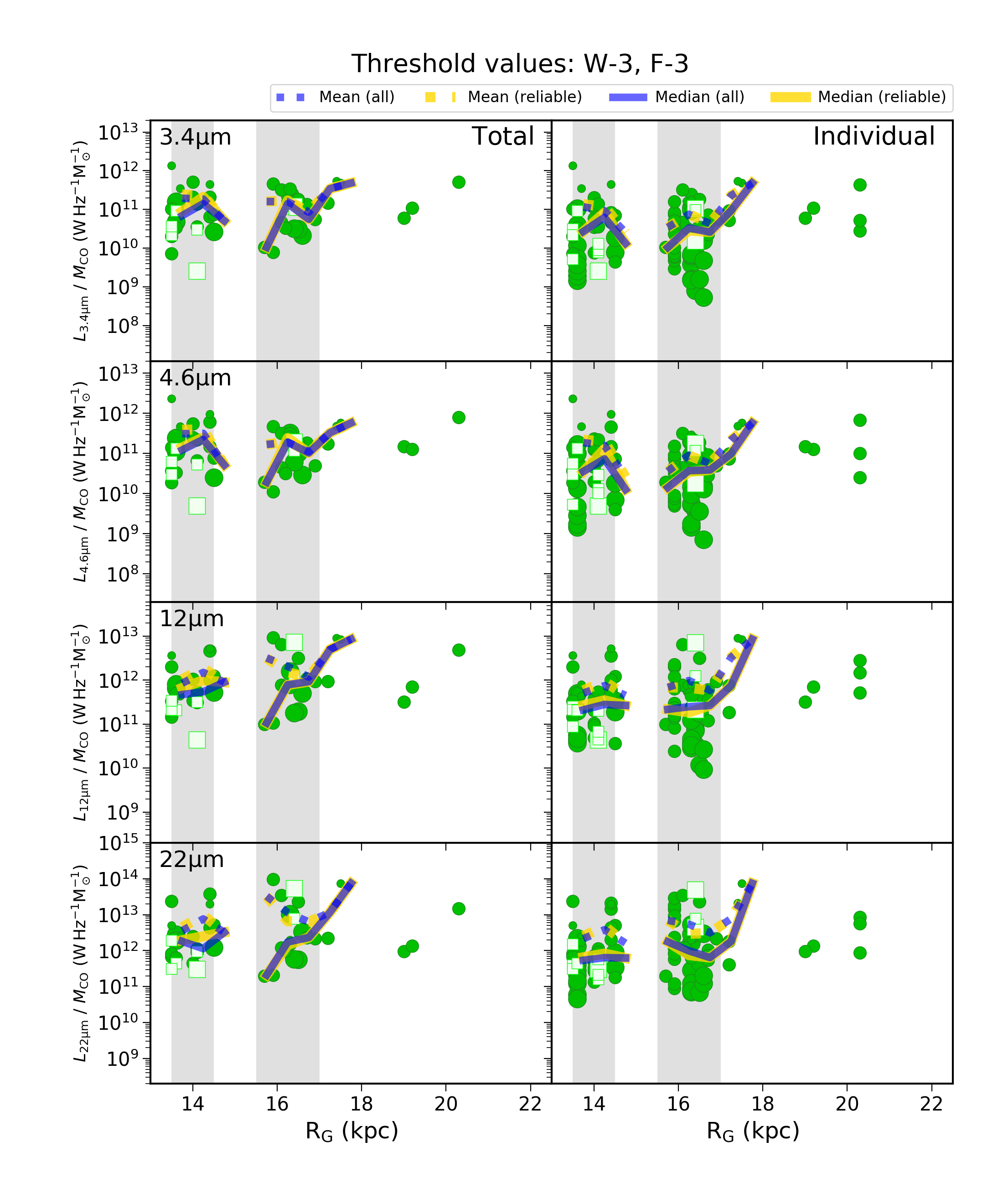}
\caption{
Identical to Figure \ref{Lumimass_CompMass-0_WISEComp-m2_rg}, but for the threshold values W-3 and F-3.}
\label{Lumimass_CompMass-2_WISEComp-m2_rg}
\end{figure*}

In this appendix, we discuss the effect of the threshold values on the two SFE-parameters: 1) $N_{\rm SF}$/$N_{\rm all}$ and 2) $L_{\rm MIR}$/$M_{\rm CO}$.
We derive the two parameters for three different threshold values in each \textit{WISE} MIR and FCRAO CO data set.
The adopted threshold values are summarized in Table 2.
Table 3 shows the results from the least-squares fittings
between $N_{\rm SF}$/$N_{\rm all}$ and $R_{\rm G}$ for $R_{\rm G}$ $\le$ 20.0 kpc.
Tables4 and 5 show the CCs between $L_{\rm MIR}$/$M_{\rm CO}$ and $R_{\rm G}$
for $R_{\rm G}$ $\le$ 20.0 kpc.
Figure \ref{Total_rate_rg_comp} shows the relation between $N_{\rm SF}$/$N_{\rm all}$ and $R_{\rm G}$ for nine combinations of the threshold values.
Figures \ref{Lumimass_CompMass-0_WISEComp-all_rg}--\ref{Lumimass_CompMass-2_WISEComp-m2_rg} show the relation between $L_{\rm MIR}$/$M_{\rm CO}$ and $R_{\rm G}$
for eight combinations of the threshold values.
Note that the relation between $L_{\rm MIR}$/$M_{\rm CO}$ and $R_{\rm G}$ of the W-2 and F-2 combination of the threshold values is shown in Figure \ref{Lumimass_rg}.
From table \ref{tab:cc_SFE1}2 and Figure \ref{Total_rate_rg_comp},
we conclude that
$N_{\rm SF}$/$N_{\rm all}$ does not decrease with increasing $R_{\rm G}$.
From tables 4, 5 and Figures \ref{Lumimass_CompMass-0_WISEComp-all_rg}--\ref{Lumimass_CompMass-2_WISEComp-m2_rg},
we confirm that $L_{\rm MIR}$/$M_{\rm CO}$ is similar at any $R_{\rm G}$ in all cases of the threshold values.
The above results are the same as those reported in Sections \ref{sec:5_21}, \ref{sec:5_22}, and \ref{sec:5_23}.
Therefore, we conclude that the different combinations of thresholds do not affect the trends in the SFE-parameters along $R_{\rm G}$.
Their actual values, nevertheless, can be different and are mostly scaling (i.e., shifting along the vertical axis in these figures) depending on the number of sources selected.

\begin{deluxetable}{cccc}
\tablecaption{Symbol names of threshold values of WISE and FCRAO data}
\tablehead{
\colhead{Symbol name} &
\colhead{Data} &
\colhead{Threshold value} &
\colhead{Note}
}
\startdata
W-1   & WISE   &  99, 99, 99, and 99 mag at 3.4, 4.6, 12, and 22 $\mu$m  & not considering any threshold value \\
W-2   & WISE   &  0.43, -0.57, -4.87, and -8.17 mag at 3.4, 4.6, 12, and 22 $\mu$m & adopted values of this paper \\
W-3   & WISE   &  -0.32, -1.32, -5.62, and -9.75 mag at 3.4, 4.6, 12, and 22 $\mu$m &  twice as the adopted values \\ \hline
F-1   & FCRAO   &  0  $M_\odot$  & not considering any threshold value \\
F-2   & FCRAO   &  183.6  $M_\odot$  & adopted value of this paper \\
F-3   & FCRAO   &  367.2  $M_\odot$  & twice as the adopted value
\enddata
\label{tab:comp_name}
\end{deluxetable}
\begin{deluxetable}{cccc}
\tablecaption{Results of the least-squares fittings between $N_{\rm SF}$/$N_{\rm all}$ and $R_{\rm G}$} 
\tablehead{
\colhead{WISE} &
\colhead{FCRAO} &
\colhead{Candidate selection\tablenotemark{a}}       &
\colhead{Results of the least-squares fitting} \\
}
\startdata
W-1 & F-1 & all                       & $N_{\rm SF}$/$N_{\rm all}$ = 3.1($\pm$1.6)$R_{\rm G}$ - 6.1($\pm$24.4) \\
    &     & only reliable ($<$ 30 \%) & $N_{\rm SF}$/$N_{\rm all}$ = 5.5($\pm$1.4)$R_{\rm G}$ - 38.1($\pm$21.3) \\
    & F-2 & all                       & $N_{\rm SF}$/$N_{\rm all}$ = 3.3($\pm$1.7)$R_{\rm G}$ - 21.1($\pm$24.9) \\
    &     & only reliable ($<$ 30 \%) & $N_{\rm SF}$/$N_{\rm all}$ = 7.6($\pm$1.3)$R_{\rm G}$ - 59.1($\pm$18.9) \\             
    & F-3 & all                       & $N_{\rm SF}$/$N_{\rm all}$ = 1.9($\pm$2.5)$R_{\rm G}$ - 50.8($\pm$37.2) \\ 
    &     & only reliable ($<$ 30 \%) & $N_{\rm SF}$/$N_{\rm all}$ = 7.0($\pm$1.8)$R_{\rm G}$ - 46.8($\pm$26.0) \\ \hline
W-2 & F-1 & all                       & $N_{\rm SF}$/$N_{\rm all}$ = 4.9($\pm$1.5)$R_{\rm G}$ - 49.5($\pm$22.4) \\
    &     & only reliable ($<$ 30 \%) & $N_{\rm SF}$/$N_{\rm all}$ = 5.2($\pm$1.3)$R_{\rm G}$ - 56.5($\pm$18.9) \\
    & F-2 & all                       & $N_{\rm SF}$/$N_{\rm all}$ = 5.8($\pm$2.5)$R_{\rm G}$ - 52.3($\pm$36.5) \\
    &     & only reliable ($<$ 30 \%) & $N_{\rm SF}$/$N_{\rm all}$ = 6.2($\pm$2.1)$R_{\rm G}$ - 64.0($\pm$31.0) \\ 
    & F-3 & all                       & $N_{\rm SF}$/$N_{\rm all}$ = 5.0($\pm$5.0)$R_{\rm G}$ - 36.7($\pm$73.4) \\ 
    &     & only reliable ($<$ 30 \%) & $N_{\rm SF}$/$N_{\rm all}$ = 5.8($\pm$4.0)$R_{\rm G}$ - 53.7($\pm$59.1) \\ \hline
W-3 & F-1 & all                       & $N_{\rm SF}$/$N_{\rm all}$ = 2.3($\pm$0.9)$R_{\rm G}$ - 23.3($\pm$13.5) \\
    &     & only reliable ($<$ 30 \%) & $N_{\rm SF}$/$N_{\rm all}$ = 2.7($\pm$0.8)$R_{\rm G}$ - 30.4($\pm$11.9) \\
    & F-2 & all                       & $N_{\rm SF}$/$N_{\rm all}$ = 3.3($\pm$1.1)$R_{\rm G}$ - 30.8($\pm$16.4) \\
    &     & only reliable ($<$ 30 \%) & $N_{\rm SF}$/$N_{\rm all}$ = 4.1($\pm$1.0)$R_{\rm G}$ - 44.7($\pm$14.5) \\             
    & F-3 & all                       & $N_{\rm SF}$/$N_{\rm all}$ = 4.3($\pm$1.6)$R_{\rm G}$ - 40.1($\pm$23.1) \\ 
    &     & only reliable ($<$ 30 \%) & $N_{\rm SF}$/$N_{\rm all}$ = 5.2($\pm$1.3)$R_{\rm G}$ - 57.1($\pm$18.3) 
\enddata
\tablenotetext{a}{The selection is based on the contamination threshold (30 \%).}
\label{tab:cc_SFE1}
\end{deluxetable}
\begin{deluxetable}{ccccccc}
\tablecaption{Correlation coefficients between $L_{\rm MIR}$/$M_{\rm CO}$ and $R_{\rm G}$ (Integrated luminosity)}
\tablehead{
\colhead{WISE} &
\colhead{FCRAO} &
\colhead{Candidate selection\tablenotemark{a}}       &
\multicolumn{4}{c}{Correlation Coefficient} \\
\cline{4-7} \colhead{} & \colhead{} & \colhead{} & \colhead{3.4 $\micron$} & \colhead{4.6 $\micron$} & \colhead{12 $\micron$} & \colhead{22 $\micron$}
}
\startdata
W-1 & F-1 & all              & -0.015 & -0.004 & -0.011 & 0.008 \\
    &     & only reliable ($<$ 30 \%) & -0.038 & -0.028 & -0.040 & -0.020 \\
    & F-2 & all              & 0.158 & 0.127 & 0.180  & 0.130  \\
    &     & only reliable ($<$ 30 \%) & 0.116 & 0.087 & 0.137 & 0.094  \\ 
    & F-3 & all              & 0.058  & 0.015  & 0.177  & 0.149 \\ 
    &     & only reliable ($<$ 30 \%) & -0.011  & -0.055  & 0.124  & 0.109 \\ \hline
W-2 & F-1 & all              & -0.075 & -0.065 & -0.079 & -0.045 \\
    &     & only reliable ($<$ 30 \%)             & -0.093 & -0.084 & -0.105 & -0.071 \\
    & F-2 & all              & 0.126 & 0.090 & 0.162  & 0.117 \\
    &     & only reliable ($<$ 30 \%)             & 0.098 & 0.063 & 0.128  & 0.086 \\
    & F-3 & all              & -0.013  & -0.070  & 0.156  & 0.140 \\ 
    &     & only reliable ($<$ 30 \%)             & -0.058  & -0.113  & 0.117  & 0.107 \\ \hline
W-3 & F-1 & all              & -0.095 & -0.075 & -0.088 & -0.049 \\
    &     & only reliable ($<$ 30 \%) & -0.129 & -0.110 & -0.137 & -0.099 \\
    & F-2 & all              & 0.220  & 0.157  & 0.264  & 0.171 \\
    &     & only reliable ($<$ 30 \%) & 0.180  & 0.117  & 0.202  & 0.114 \\ 
    & F-3 & all              & -0.025  & -0.113  & 0.255  & 0.208 \\ 
    &     & only reliable ($<$ 30 \%)  & -0.093  & -0.181  & 0.182  & 0.145 
\enddata
\tablenotetext{a}{The selection is based on the contamination threshold (30 \%).}
\label{tab:cc_SFE2-total}
\end{deluxetable}
\begin{deluxetable}{ccccccc}
\tablecaption{Correlation coefficients between $L_{\rm MIR}$/$M_{\rm CO}$ and $R_{\rm G}$ (Individual luninosity)}
\tablehead{
\colhead{WISE} &
\colhead{FCRAO} &
\colhead{Candidate selection\tablenotemark{a}}       &
\multicolumn{4}{c}{Correlation Coefficient} \\
\cline{4-7} \colhead{} & \colhead{} & \colhead{} & \colhead{3.4 $\micron$} & \colhead{4.6 $\micron$} & \colhead{12 $\micron$} & \colhead{22 $\micron$}
}
\startdata
W-1 & F-1 & all              & 0.001 & 0.009 & 0.005 & 0.015 \\
    &     & only reliable ($<$ 30 \%) & -0.013 & -0.005 & -0.013 & -0.003\\
    & F-2 & all              & 0.143 & 0.117 & 0.149  & 0.122 \\
    &     & only reliable ($<$ 30 \%) & 0.119 & 0.094 & 0.127  & 0.101 \\             
    & F-3 & all              & 0.055  & 0.025  & 0.119  & 0.117 \\ 
    &     & only reliable ($<$ 30 \%) & 0.020  & -0.007  & 0.092  & 0.093 \\ \hline
W-2 & F-1 & all              & -0.041 & -0.032 & -0.041 & -0.020 \\
    &     & only reliable ($<$ 30 \%)             & -0.053 & -0.045 & -0.059 & -0.039 \\
    & F-2 & all              & 0.136 & 0.101 & 0.158 & 0.131 \\
    &     & only reliable ($<$ 30 \%)             & 0.116 & 0.083 & 0.136 & 0.109\\ 
    & F-3 & all              & -0.002  & -0.043  & 0.122  & 0.130 \\ 
    &     & only reliable ($<$ 30 \%)             & -0.034 & -0.071 & 0.094  & 0.104 \\ \hline
W-3 & F-1 & all              & -0.072 & -0.055  & -0.064 & -0.036 \\
    &     & only reliable ($<$ 30 \%) & -0.093 & -0.077 & -0.097 & -0.071 \\
    & F-2 & all              & 0.200  & 0.145  & 0.240  & 0.188  \\
    &     & only reliable ($<$ 30 \%) & 0.179  & 0.123  & 0.204  & 0.147 \\ 
    & F-3 & all              & -0.024 & -0.089  & 0.203 & 0.204  \\ 
    &     & only reliable ($<$ 30 \%) & -0.062 & -0.127 & 0.157 & 0.156
\enddata
\tablenotetext{a}{The selection is based on the contamination threshold (30 \%).}
\label{tab:cc_SFE2-ind}
\end{deluxetable}
\clearpage
\section{Distributions}\label{sec:a_2}
\begin{figure*}
\epsscale{1.0}
\plotone{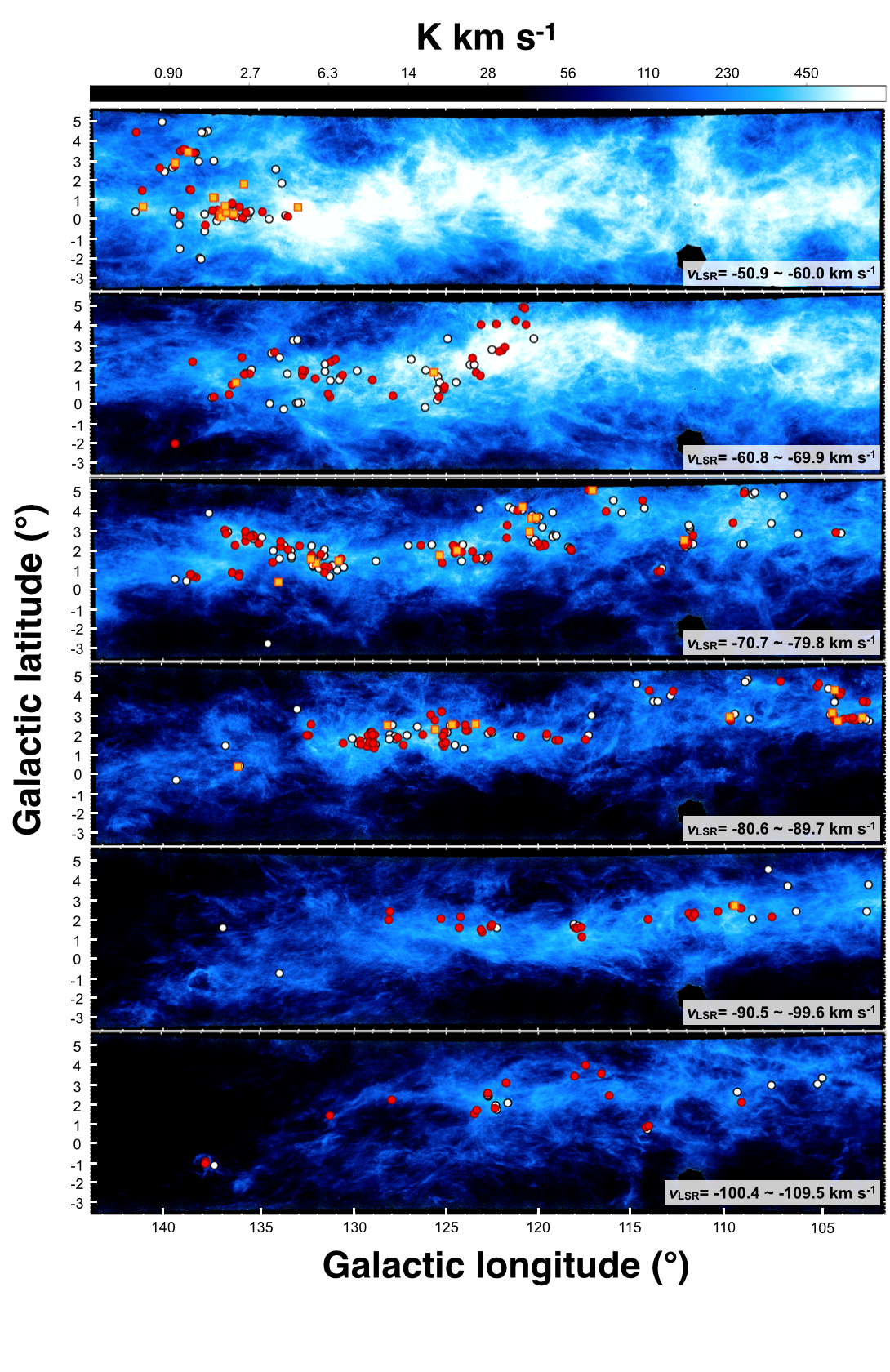}
\caption{
Locations of BKP clouds overplotted on \ion{H}{1} channel maps from 
the Canadian Galactic Plane Survey \citep{Taylor2003}.
Black open circles show clouds without associated star-forming regions.
Red circles and yellow squares show clouds with associated candidate star-forming regions with a contamination rate of their parental clouds
of $<$ 30 \% (reliable) and $\ge$ 30 \% (less reliable), respectively.
} 
\label{BKP-HI_lb}
\end{figure*}

In this appendix, we report the distribution of BKP clouds.
Figure \ref{BKP-HI_lb} displays the locations of BKP clouds on the \ion{H}{1} $l$-$b$ channel maps in the range
-109.5 km s$^{-1}$ $\le$ $v_{\rm LSR}$ $\le$ -50.9 km s$^{-1}$.
\ion{H}{1} clouds form shells and filamentary structures,
and almost all BKP clouds are associated with those \ion{H}{1} structures.
Many BKP clouds are located at $b$ $\ge$ 1$^\circ$, which is expected from the Galactic warping
\citep[e.g.,][]{Nakanishi2016}.
As in Figure \ref{BKP-HI_lb}, almost all BKP clouds are associated with bright \ion{H}{1} components.
We could not find any clear difference between the distribution of BKP clouds with and without candidate star-forming regions.


\clearpage
\bibliographystyle{aasjournal}

\bibliography{Ref}



\end{document}